\documentclass[graybox]{svmult}

\usepackage{mathptmx}       
\usepackage{helvet}         
\usepackage{courier}        
\usepackage{type1cm}        
\usepackage{amssymb}
\usepackage{amsmath}
\usepackage{amsfonts}
\usepackage{hyperref}
\usepackage{pictexwd,dcpic}
\usepackage{makeidx}         
\usepackage{graphicx}        
\usepackage{multicol}        
\usepackage[bottom]{footmisc}

\makeindex             

\def\half{{1\over2}}

\def\g{\gamma}

\newcommand{\meV}{{\rm meV}}
\newcommand{\eV}{{\rm eV}}
\newcommand{\keV}{{\rm keV}}
\newcommand{\MeV}{{\rm MeV}}
\newcommand{\GeV}{{\rm GeV}}
\newcommand{\TeV}{{\rm TeV}}

\newcommand{\kpc}{{\rm kpc}}

\newcommand{\s}{{\rm s}}

\newcommand{\Mpl}{M_{\rm Pl}}

\newcommand{\im}{\mathrm{i}}

\newcommand{\bnabla}{\mathbf{\nabla}}

\def\csound{c_{\rm sound}}
\def\Vext{V_{\rm ext}}
\newcommand{\eeq}{\end{equation}} 
\newcommand{\beq}{\begin{equation}} \newcommand{\ee}{\end{equation}} \newcommand{\eq}{\end{equation}}
\def\x{{\bf x}}

\begin{document}

\title*{Lorentz breaking Effective Field Theory and observational tests}
\author{Stefano Liberati}
\institute{Stefano Liberati \at SISSA, Via Bonomea 265, 34136, Trieste, Italy\\ \email{liberati@sissa.it}
}
%
%
\maketitle

\abstract*{Analogue models of gravity have provided an experimentally realizable test field for our ideas on quantum field theory in curved spacetimes but they have also inspired the investigation of possible departures from exact Lorentz invariance at microscopic scales. In this role they have joined, and sometime anticipated, several  quantum gravity models characterized by Lorentz breaking phenomenology. A crucial difference between these speculations and other ones associated to quantum gravity scenarios, is the possibility to carry out observational and experimental tests which have nowadays led to a broad range of constraints on departures from Lorentz invariance. We shall review here the effective field theory approach to Lorentz breaking in the matter sector, present the constraints provided by the available observations and finally discuss the implications of the persisting uncertainty on the composition of the ultra high energy cosmic rays for the constraints on the higher order, analogue gravity inspired, Lorentz violations.}

\abstract{Analogue models of gravity have provided an experimentally realizable test field for our ideas on quantum field theory in curved spacetimes but they have also inspired the investigation of possible departures from exact Lorentz invariance\index{Lorentz invariance} at microscopic scales. In this role they have joined, and sometime anticipated, several  quantum gravity models characterized by Lorentz breaking phenomenology. A crucial difference between these speculations and other ones associated to quantum gravity scenarios, is the possibility to carry out observational and experimental tests which have nowadays led to a broad range of constraints on departures from Lorentz invariance. We shall review here the effective field theory approach to Lorentz breaking in the matter sector, present the constraints provided by the available observations and finally discuss the implications of the persisting uncertainty on the composition of the ultra high energy cosmic rays\index{high energy cosmic rays} for the constraints on the higher order, analogue gravity inspired, Lorentz violations\index{Lorentz violations}.}

\section{Introduction}
\label{sec:1}

Our understanding of the fundamental laws of Nature is based at present on two different theories: the Standard Model of Fundamental Interactions (SM), and classical General Relativity (GR).  However, in spite of their phenomenological success, SM and GR leave many theoretical questions still unanswered. First of all, since we feel that our understanding of the fundamental laws of Nature is deeper (and more accomplished) if we are able to reduce the number of degrees of freedom and coupling constants we need in order to describe it, many physicists have been trying to construct unified theories in which not only sub-nuclear forces are seen as different aspects of a unique interaction, but also gravity is included in a consistent manner.  

Another important reason why we seek for a new theory of gravity comes directly from the gravity side. We know that GR fails to be a predictive theory in some regimes. Indeed, some solutions of Einstein's equations are known to be singular at some points, meaning that in these points GR is not able to make any prediction. Moreover, there are apparently honest solutions of GR equations predicting the existence of time-like closed curves, which would imply the possibility of traveling back and forth in time with the related causality paradoxes. Finally, the problem of black-hole evaporation seems to clash with Quantum Mechanical unitary evolution. 

This long list of puzzles spurred an intense research toward a quantum theory of gravity that started almost immediately after Einstein's proposal of GR and which is still going on nowadays. The quantum gravity problem is not only conceptually challenging, it has also been an almost metaphysical pursue for several decades. Indeed, we expect QG effects at experimentally/observationally accessible energies to be extremely small, due to suppression by the Planck scale $M_{\rm pl} \equiv \sqrt{\hbar c/G_{\rm N}}\simeq 1.22\times 10^{19}~\mbox{GeV}/c^{2}$. In this sense it has been considered (and it is still considered by many) that only ultra-high-precision (or Planck scale energy) experiments would be able to test quantum gravity models.

It was however realized (mainly over the course of the past decade) that the situation is not as bleak as it appears. In fact, models of gravitation beyond GR and models of QG have shown that there can be several of what we term low energy ``relic signatures'' of these models, which would lead to deviation from the standard theory predictions (SM plus GR) in specific regimes. Some of these new phenomena, which comprise what is often termed ``QG phenomenology'', include: 
\begin{itemize}
\item Quantum decoherence and state collapse \cite{Mavromatos:2004sz}
\item QG imprint on initial cosmological perturbations \cite{Weinberg:2005vy}
\item Cosmological variation of couplings \cite{Damour:1994zq,Barrow:1997qh}
\item TeV Black Holes, related to extra-dimensions \cite{Bleicher:2001kh}
\item Violation of discrete symmetries \cite{Kostelecky:2003fs}
\item Violation of space-time symmetries \cite{Mattingly:2005re}
\end{itemize}
In this lecture I will focus upon the phenomenology of violations of space-time symmetries, and in particular of Local Lorentz invariance (LLI), a pillar both of quantum field theory as well as GR (LLI is a crucial part of the Einstein Equivalence Principle on which metric theories of gravity are based). 

\section{A brief history of an heresy}
\label{sec:LVhistory}

Contrary to the common trust, ideas about the possible breakdown of LLI have a long standing history. It is however undeniable that the last twenty years have witnessed a striking acceleration in the development both of theoretical ideas as well as of phenomenological tests before unconceivable. We shall here present an incomplete review of these developments.

\subsection{The dark ages}

The possibility that Lorentz invariance violation (LV) could play a role again in physics dates back by at least sixty years~\cite{DiracAet,Bj,Phillips66,Blokh66,Pavl67,Redei67} and in the seventies and eighties there was already a well established  literature investigating the possible phenomenological consequences of LV (see e.g.~\cite{1978NuPhB.141..153N,1980NuPhB.176...61E,Zee:1981sy,Nielsen:1982kx,1983NuPhB.217..125C,1983NuPhB.211..269N}).  

The relative scarcity of these studies in the field was due to the general expectation that new effects were only to appear in particle interactions at energies of order the Planck mass $M_{\rm pl}$.  However, it was only in the nineties that it was clearly realized that there are special situations in which new effects could manifest also at lower energy. These situations were termed ``Windows on Quantum Gravity''.

\subsection{Windows on Quantum Gravity\index{Windows on Quantum Gravity}}

At first glance, it appears hopeless to search for effects suppressed by the Planck scale. Even the most energetic particles ever detected (Ultra High Energy Cosmic Rays, see, e.g.,~\cite{Roth:2007in,Abbasi:2007sv}) have $E \lesssim 10^{11}$ GeV $\sim 10^{-8} M_{\mathrm pl}$. 
However, even tiny corrections can be magnified into a significant effect when dealing with high energies (but still well below the Planck scale), long distances of signal propagation, or peculiar reactions (see, e.g.,~\cite{Mattingly:2005re} for an extensive review).

A partial list of these {\em windows on QG} includes:
\begin{itemize}
\item sidereal variation of LV couplings as the lab moves
  with respect to a preferred frame or direction
\item cumulative effects: long baseline dispersion and vacuum birefringence (e.g.~of signals from gamma ray bursts, active galactic nuclei, pulsars)
\item anomalous (normally forbidden) threshold reactions allowed by LV terms (e.g.~photon decay, vacuum Cherenkov effect) 
\item shifting of existing threshold reactions (e.g.~photon annihilation from Blazars, ultra high energy protons pion production)
\item LV induced decays not characterized by a threshold (e.g.~decay of a particle from one helicity to the other or photon splitting)
\item maximum velocity (e.g.~synchrotron peak from supernova remnants)
\item dynamical effects of LV background fields (e.g. gravitational coupling and additional wave modes)
\end{itemize}

It is difficult to assign a definitive ``paternity" to a field, and the so called Quantum Gravity Phenomenology\index{Quantum Gravity Phenomenology} is no exception in this sense.  However, among the papers commonly accepted as seminal we can cite  the one by Kosteleck\'{y} and Samuel \cite{KS89} that already in 1989 envisaged, within a string field theory framework, the possibility of non-zero vacuum expectation values (VEV) for some Lorentz breaking operators. This work led later on to the development of a systematic extension of the SM (what was later on called "minimal standard model extension" (mSME)) incorporating all possible Lorentz breaking, power counting renormalizable, operators (i.e. of mass dimension $\leq 4$), as proposed by Colladay and Kosteleck\'{y}~\cite{Colladay:1998fq}.  This provided a framework for computing in effective field theory the observable consequences for many experiments and led to much experimental work setting limits on the LV parameters in the Lagrangian (see e.g.~\cite{Kostelecky:2008zz} for a review).
 
Another seminal paper was that of Amelino-Camelia and collaborators~\cite{AmelinoCamelia:1997gz} which highlighted the possibility to cast observational constraints on high energy violations of Lorentz invariance in the photon dispersion relation by using the aforementioned propagation over cosmological distance of light from remote astrophysical sources like gamma ray bursters (GRBs) and active galactic nuclei (AGN). The field of phenomenological constraints on quantum gravity induced LV was born.

Finally, we should also mention the influential papers by Coleman and Glashow \cite{Coleman:1997xq,Coleman:1998en,Coleman:1998ti} which brought the subject of systematic tests of Lorentz violation to the attention of of the broader community of particle physicists. 

Let me stress that this is necessarily an incomplete account of the literature which somehow pointed a spotlight on the investigation of departures from Special Relativity. Several papers appeared in the same period and some of them anticipated many important results, see e.g.\cite{GonzalezMestres:1996zv,GonzalezMestres:1997if}, which however at the time of their appearance were hardly noticed (and seen by many as too ``exotic").

In the years 2000 the field reached a concrete maturity and many papers pursued a systematization both of the framework as well as of the available constraints (see e.g.~\cite{Jacobson:2002hd, Mattingly:2002ba, Jacobson:2005bg}). In this sense another crucial contribution was the development of an effective field theory approach also for higher order (mass dimension greater than four), naively non-power counting renormalizable, operators \footnote{Anisotropic scaling \cite{Anselmi:2008ry,Horava:2009uw,Visser:2009ys} techniques were recently recognized to be the most appropriate way of handling higher order operators in Lorentz breaking theories and in this case the highest order operators are indeed crucial in making the theory power counting renormalizable. This is why we shall adopt sometime the expression ``naively non renormalizable"}. This was firstly done for dimension 5 operators in QED \cite{Myers:2003fd} by Myers and Pospelov and later on extended to dimension 6 operators by Mattingly \cite{Mattingly:2008pw}.

Why all this attention to Lorentz breaking tests developed in the late nineties and in the first decade of the new century? I think that the answer is twofold as it is related to important developments coming from experiments and observation as well as from theoretical investigations.
It is a fact that the zoo of quantum gravity models/scenarios with a low energy phenomenology had a rapid growth during those years. This happened mainly under the powerful push of novel puzzling observations that seemed to call for new physics possibly of gravitational origin. For example, in cosmology these are the years of the striking realization that our universe is undergoing an accelerated expansion phase \cite{Riess:1998cb,Perlmutter:1998np} which apparently requires a new exotic cosmological fluid, called dark energy, which violates the strong energy condition (to be added to the already well known, and still mysterious, dark matter component).

 Also in the same period high energy astrophysics provided some new puzzles, first with the apparent absence of the Greisen-Zatsepin-Kuzmin (GZK) cut off \cite{Greisen:1966jv,1969cora...11...45Z} (a suppression of the high-energy tail of the UHECR spectrum due to UHECR interaction with CMB photons) as claimed by the Japanese experiment AGASA \cite{Takeda:1998ps}, later on via the so called TeV-gamma rays crisis, i.e. the apparent detection of a reduced absorption of TeV gamma rays emitted by AGN \cite{Protheroe:2000hp}. Both these ``crises" later on subsided or at least alternative, more orthodox,  explanations for them were advanced. However, their undoubtedly boosted the research in the field at that time.

It is perhaps this past ``training" that made several exponents of the quantum gravity phenomenology community the among most ready to stress the apparent incompatibility of the recent CERN--LNGS based experiment OPERA~\cite{Opera:2011zb} measure of superluminal propagation of muonic neutrinos and Lorentz EFT (see e.g.~\cite{AmelinoCamelia:2011dx,Cohen:2011hx,Maccione:2011fr,Carmona:2011zg}. There is now evidence that the Opera measurement might be flawed due to unaccounted experimental errors and furthermore it seems to be refuted by a similar measurement of the ICARUS collaboration~\cite{Antonello:2012hg}.  Nonetheless, this claim propelled a new burst of activity in Lorentz breaking phenomenology which might still provide useful insights for future searches.

Parallel to these exciting developments on the experimental/observational side, also theoretical investigations provided new motivations for Lorentz breaking searches and constraints. Indeed, specific hints of LV arose from various approaches to Quantum Gravity. Among the many examples are the above mentioned string theory tensor VEVs \cite{KS89} and space-time foam models~\cite{AmelinoCamelia:1996pj,AmelinoCamelia:1997gz,Ellis:1999jf,Ellis:2000sx,Ellis:2003sd}, then semiclassical spin-network calculations in Loop QG~\cite{Gambini:1998it}, non-commutative geometry~\cite{Carroll:2001ws, Lukierski:1993wx, AmelinoCamelia:1999pm}, some brane-world backgrounds~\cite{Burgess:2002tb}. 

Indeed, during the last decades there were several attempts to formulate alternative theories of gravitation incorporating some form of Lorentz breaking, from early studies~\cite{Gasperini:1985aw,Gasperini:1986xb,Gasperini:1987nq,Gasperini:1987fq,Gasperini:1998eb} to full-fledged theories such as the Einstein--Aether theory \cite{Mattingly:2001yd,Eling:2004dk,Jacobson:2008aj} and Ho\v rava--Lifshitz  gravity \cite{Horava:2009uw,Sotiriou:2009bx,Blas:2009qj} (which in some limit can be seen as a UV completion of the Einstein--Aether framework~\cite{Jacobson:2010mx}). 

Finally, a relevant part of this story is related to the vigorous development in the same years of the so called condensed matter analogues of ``emergent gravity''~\cite{Barcelo:2005fc} which is the main topic of this school. Let us then consider these models in some detail and discuss some lesson that can be drawn from them.

\section{Bose--Einstein condensates as an example of emergent Local Lorentz invariance}

Analogue models for gravity\index{Analogue models for gravity} have provided a powerful tool for testing (at least in principle) kinematical features of classical and quantum field theories in curved spacetimes \cite{Barcelo:2005fc}. The typical setting is the one of sound waves propagating in a perfect fluid \cite{Unruh:1980cg,Visser:1993ub}. Under certain conditions, their equation can be put in the form of a Klein-Gordon equation for a massless particle in curved spacetime, whose geometry is specified by the acoustic metric. Among the various condensed matter systems so far considered, Bose--Einstein condensate\index{Bose--Einstein condensate} (BEC) \cite{Garay:1999sk,Barcelo:2000tg} had in recent years a prominent role for their simplicity as well as for the high degree of sophistication achieved by current experiments. In a BEC system one can consider explicitly the quantum field theory of the quasi-particles (or phonons), the massless excitations over the condensate state, propagating over the condensate as the analogue of a quantum field theory of a scalar field propagating over a curved effective spacetime described by the acoustic metric. It provides therefore a natural framework to explore different aspects of quantum field theory in various interesting curved backgrounds (for example quantum aspects of black hole physics \cite{Balbinot:2004da,Barcelo:2007yk} or the analogue of the creation of cosmological perturbations \cite{Barcelo:2003wu, 
Weinfurtner:2008ns,Weinfurtner:2008if,Jain:2007gg}) or even, and more relevantly for our discussion here, emerging spacetime scenarios.

In BEC, the effective emerging metric depends on the properties of the condensate wave-function. One can expect therefore the gravitational degrees of freedom to be encoded in the variables describing the condensate wave-function \cite{Barcelo:2000tg}, which is solution of the well known Bogoliubov--de Gennes (BdG) equation. The dynamics of gravitational degrees of freedom should then be  inferred from this  equation, which is essentially non-relativistic.   The ``emerging matter", the quasi-particles, in the standard BEC, are phonons, i.e.~massless excitations described at low energies by a relativistic (we shall see in which sense) wave equation, however, at high energies, the emergent nature of the underlying spacetime becomes evident and  the relativistic structure of the equation broken. Let's see this in more detail as a conceptual exercise and for highlighting the inspirational role played in this sense by analogue models of gravity.

\subsection{The acoustic geometry in BEC}
Let us start by very briefly reviewing the derivation of the acoustic
metric for a BEC system, and show that the equations for the phonons
of the condensate closely mimic the dynamics of a scalar field in a
curved spacetime.  In the dilute gas approximation, one can describe a
Bose gas through a quantum field ${\widehat \Psi}$ satisfying
\begin{eqnarray}
\im\hbar \; \frac{\partial }{\partial t} {\widehat \Psi} =
\left(
  - \frac{\hbar^2}{2m} \nabla^2 + \Vext(\x)
  +\kappa(a)\;{\widehat \Psi}^{\dagger}{\widehat \Psi}
\right){\widehat \Psi}. \label{eq:fieldeq0}
\end{eqnarray}
$m$ is the mass of the atoms,  $a$ is the scattering length for the atoms and $\kappa$ parametrises the strength of the interactions between the different bosons in the gas.
It can be re-expressed in terms of the scattering length $a$ as
\begin{equation}
\kappa(a) = \frac{4\pi a \hbar^2}{m}.
\end{equation}
As usual, the quantum field can be separated into
a macroscopic (classical) condensate and a fluctuation: ${\widehat
\Psi}=\psi+{\widehat \varphi}$, with $\langle {\widehat \Psi}
\rangle=\psi $. Then, by adopting the self-consistent mean field
approximation
%
\begin{eqnarray}
{\widehat \varphi}^{\dagger}{\widehat \varphi}{\widehat \varphi}
\simeq
2\langle {\widehat \varphi}^{\dagger}{\widehat \varphi} \rangle \;
{\widehat \varphi}
+ \langle {\widehat \varphi} {\widehat \varphi}
\rangle \; {\widehat \varphi}^{\dagger},
\end{eqnarray}
one can arrive
at the set of coupled equations:
\begin{eqnarray}
\im \hbar\;\frac{\partial }{\partial t} \psi(t,\x)
&=&
\left ( - {\hbar^2 \over
2m} \nabla^2 + \Vext(\x) + \kappa \; n_c \right) \psi(t,\x)
\nonumber\\
&&
\qquad + \kappa \left\{2\tilde n  \psi(t,\x)+ \tilde m \psi^*(t,\x) \right\};
\label{bec-self-consistent1}
\\
&& \nonumber\\
\im \hbar \; \frac{\partial }{\partial t} {\widehat \varphi}(t,\x)
&=&
\left( - {\hbar^2 \over 2m} \nabla^2  +
\Vext(\x)   +\kappa \;2 n_T \right){\widehat \varphi} (t,\x)
\nonumber
\\
&&
\qquad +  \kappa \; m_T \; {\widehat \varphi}^{\dagger}(t,\x).
\label{quantum-field}
\end{eqnarray}
Here
\begin{eqnarray}
&& n_c \equiv \left| \psi(t,\x) \right|^2;
\quad
m_c \equiv \psi^2(t,\x);
\\
&& \tilde n \equiv  \langle
{\widehat \varphi}^{\dagger}\,{\widehat \varphi} \rangle;
\quad \quad  \quad
\tilde m \equiv  \langle {\widehat \varphi}\, {\widehat \varphi} \rangle;
\\
&& n_T=n_c+\tilde n;
\quad \quad
m_T=m_c+\tilde m.
\end{eqnarray}
In general one will have to solve both equations for $\psi$ and $\widehat \phi$ simultaneously. The equation for the condensate wave function $\psi$ is
closed only when the back-reaction effects due to the fluctuations are neglected. (The back-reaction being hidden in the quantities $\tilde m$ and $\tilde n$.) This  approximation leads then to the so-called Gross--Pitaevskii equation and can be checked {\em a posteriori} to be a good description of dilute Bose--Einstein condensates near equilibrium configurations.

Adopting the Madelung representation for
the wave function $\psi$ of the condensate
\begin{equation}
\psi(t,\x)=\sqrt{n_c(t,\x)} \; \exp[-\im\theta(t,\x)/\hbar],
\end{equation}
and defining an irrotational ``velocity field'' by $\mathbf{v}\equiv{\bnabla\theta}/{m}$, the Gross--Pitaevskii equation can be rewritten
as a continuity equation plus an Euler equation:
\begin{eqnarray}
&& \frac{\partial}{\partial t}n_c+\bnabla\cdot({n_c \mathbf{v}})=0,
\label{E:continuity}\\
&& m\frac{\partial}{\partial t}\mathbf{v}+\bnabla\left(\frac{mv^2}{2}+
V_\mathrm{ext}(t,\x)+\kappa n_c- \frac{\hbar^2}{2m}
\frac{\nabla^{2}\left(\sqrt{n_c}\right)}{\sqrt{n_c}} \right)=0.
\label{E:Euler1}
\end{eqnarray}
These equations are completely equivalent to those of an irrotational
and inviscid fluid apart from the existence of the so-called quantum
potential
\begin{equation}
V_\mathrm{quantum}=
-\hbar^2\nabla^{2}\sqrt{n_c}/(2m\sqrt{n_c}),
\end{equation}
which has the dimensions of an energy. 

If we write the mass density of the Madelung fluid as $\rho
= m \; n_c$, and use the fact that the flow is irrotational 
we can write the Euler equation in the more convenient Hamilton--Jacobi form:
\begin{equation} m
\frac{\partial}{\partial t}\theta+ \left( \frac{[\bnabla\theta]^2}{2m}
+V_\mathrm{ext}(t,\x)+\kappa n_c-
\frac{\hbar^2}{2m}\frac{\nabla^{2}\sqrt{n_c}}{\sqrt{n_c}}
\right)=0. \label{E:HJ} \end{equation}
When the gradients in the
density of the condensate are small one can neglect the quantum stress
term leading to the standard hydrodynamic approximation.


Let us now consider the quantum perturbations above the condensate. These can be described in
several different ways, here we are interested in the ``quantum
acoustic representation''
\begin{eqnarray}
\widehat
\varphi(t,\x)=e^{-\im \theta/\hbar} \left({1 \over 2
\sqrt{n_c}} \; \widehat n_1 - \im \; {\sqrt{n_c} \over \hbar} \;\widehat
\theta_1\right),
\label{representation-change}
\end{eqnarray}
where
$\widehat n_1,\widehat\theta_1$ are real quantum fields.  By using
this representation Equation~(\ref{quantum-field}) can be rewritten as
\begin{eqnarray}
&&\partial_t \widehat n_1 + {1\over m}
\bnabla\cdot\left( n_1 \; \bnabla \theta + n_c \; \bnabla \widehat
\theta_1 \right) = 0, \label{pt1}
\\ &&\partial_t \widehat \theta_1   +
{1\over m} \bnabla \theta \cdot \bnabla \widehat \theta_1
+ \kappa(a) \; n_1 - {\hbar^2\over2 m}\; D_2 \widehat n_1 = 0.
\label{pt2}
\end{eqnarray}
Here $D_2$ represents a second-order differential operator obtained
from linearizing the quantum potential. Explicitly:
\begin{eqnarray}
D_2\, \widehat n_1 &\equiv&
-\half n_c^{-3/2} \;[\nabla^2
(n_c^{+1/2})]\; \widehat n_1
+\half n_c^{-1/2} \;\nabla^2
(n_c^{-1/2}\; \widehat n_1).
\end{eqnarray}
The equations we have just written can be obtained easily by
linearizing the Gross--Pitaevskii equation around a classical
solution: $n_c \rightarrow n_c + \widehat n_1$, $\phi \rightarrow \phi
+ \widehat \phi_1$.  It is important to realize that in those
equations the back-reaction of the quantum fluctuations on the
background solution has been assumed negligible.  We also see in
Equations~(\ref{pt1}, \ref{pt2}), that time variations of
$V_\mathrm{ext}$ and time variations of the scattering length $a$
appear to act in very different ways.  Whereas the external potential
only influences the background Equation~(\ref{E:HJ}) (and hence the
acoustic metric in the analogue description), the scattering length
directly influences both the perturbation and background equations.
{From} the previous equations for the linearised perturbations it is
possible to derive a wave equation for $\widehat \theta_{1}$ (or
alternatively, for $\widehat n_{1}$). All we need is to substitute in
Equation~(\ref{pt1}) the $\widehat n_{1}$ obtained from
Equation~(\ref{pt2}).  This leads to a PDE that is second-order in
time derivatives but infinite order in space derivatives -- to
simplify things we can construct the symmetric $4 \times 4$ matrix
\begin{equation}
f^{\mu\nu}(t,\x) \equiv
\begin{bmatrix}
   f^{00}&\vdots&f^{0j}\\
   \cdots\cdots&\cdot&\cdots\cdots\cdots\cdots\cr
   f^{i0}&\vdots&f^{ij}\\
\end{bmatrix}.
\label{E:explicit}
\end{equation}
(Greek indices run from
$0$--$3$, while Roman indices run from $1$--$3$.)  Then, introducing
(3+1)-dimensional space-time coordinates
\begin{equation}
x^\mu \equiv (t;\, x^i)
\end{equation}
the wave equation for $\theta_{1}$ is easily rewritten as
\begin{equation}
\partial_\mu ( f^{\mu\nu} \;
\partial_\nu \widehat \theta_1) = 0. \label{weq-phys}
\end{equation}
Where the $f^{\mu\nu}$ are \emph{differential operators} acting on space
only.
Now, if we make a spectral decomposition of the field $\widehat
\theta_1$ we can see that for wavelengths larger than $\xi=\hbar /m\csound$
($\xi$ corresponds to the ``healing length'' and $\csound(a,n_c)^2={\kappa(a) \; n_c \over m}$), the terms coming from the linearization of the quantum
potential (the $D_2$) can be neglected in the previous expressions, in
which case the $f^{\mu\nu}$ can be approximated by scalars, instead of
differential operators. Then, by identifying
\begin{equation}
\sqrt{-g} \; g^{\mu\nu}=f^{\mu\nu},
\end{equation}
the equation for the field $\widehat \theta_1$
becomes that of a (massless minimally coupled) quantum scalar field
over a curved background
\begin{equation}
\Delta\theta_{1}\equiv\frac{1}{\sqrt{-g}}\;
\partial_{\mu}\left(\sqrt{-g}\; g^{\mu\nu}\; \partial_{\nu}\right)
\widehat\theta_{1}=0,
\end{equation}
with an effective metric of the form
\begin{equation} g_{\mu\nu}(t,\x) \equiv {n_c\over m\;
\csound(a,n_c)}
\begin{bmatrix}
   -\{\csound(a,n_c)^2-v^2\}&\vdots& - v_j \\
   \cdots\cdots\cdots\cdots&\cdot&\cdots\cdots\\
   -v_i&\vdots&\delta_{ij}\\
\end{bmatrix}.
\end{equation}
Here the magnitude
$\csound(n_c,a)$ represents the speed of the phonons in the medium:
\begin{equation}
\csound(a,n_c)^2={\kappa(a) \; n_c \over m},
\end{equation}
{and $v_i$ is the velocity field of the fluid flow,}
\begin{equation}
v_i= \frac{1}{m}\nabla_i \theta.
\end{equation}

\subsection{Lorentz violation in BEC}

It is interesting to consider the case in which the above
``hydrodynamical" approximation for BECs does not hold. In order to
explore a regime where the contribution of the quantum potential
cannot be neglected we can use the so called {\emph{eikonal}}
approximation, a high-momentum approximation where the phase
fluctuation $\widehat \theta_1$ is itself treated as a slowly-varying
amplitude times a rapidly varying phase. This phase will be taken to
be the same for both $\widehat n_1$ and $\widehat \theta_1$
fluctuations. In fact, if one discards the unphysical possibility that
the respective phases differ by a time varying quantity, {any
time-independent difference can be safely reabsorbed in the definition of
the (complex) amplitudes $\mathcal{A}_\theta,\, \mathcal{A}_\rho$.}  Specifically, we shall write
\begin{eqnarray}
{\widehat\theta}_1(t,\x)
&=& \mathrm{Re}\left\{\mathcal{A}_\theta \; \exp(-i\phi) \right\},\\
{\widehat n}_1(t,\x)
&=& \mathrm{Re}\left\{\mathcal{A}_\rho \; \exp(-i\phi)\right\}.
\end{eqnarray}
As a consequence of our starting assumptions, gradients of the
amplitude, and gradients of the background fields, are systematically
ignored relative to gradients of $\phi$.  Note however, that what we are doing
here is not quite a ``standard'' eikonal approximation, in the sense
that it is not applied directly on the fluctuations of the field
$\psi(t,\x)$ but separately on their amplitudes and phases $\rho_{1}$
and $\phi_{1}$.  We can then adopt the notation
\begin{equation}
\omega ={\partial\phi\over\partial t};\qquad k_i = \nabla_i \phi.
\end{equation}
Then the operator $D_2$ can be approximated as
\begin{eqnarray}
D_2 \;{\widehat n}_1
&\approx&
-\half n_c^{-1}
\;k^2 \;{\widehat n}_1.
\end{eqnarray}
A similar result holds for
$D_2$ acting on ${\widehat \theta}_1$.  That is, under the eikonal
approximation we effectively replace the {\emph{operator}} $D_2$ by
the {\emph{function}}
\begin{equation}
D_2 \to -\half n_c^{-1}k^2.
\end{equation}
For the matrix $f^{\mu\nu}$ this effectively
results in the replacement
\begin{eqnarray}
f^{00}
&\to& - \left[\kappa(a) + {\hbar^2 \; k^2\over4m\;n_c} \right]^{-1} \\
f^{0j}
&\to& -\left[ \kappa(a) + {\hbar^2 \; k^2\over4m\;n_c}\right]^{-1}\;
{\nabla^j \theta_0\over m} \\
f^{i0}
&\to& - {\nabla^i \theta_0\over m} \;
\left[ \kappa(a) + {\hbar^2 \; k^2\over4m\;n_c} \right]^{-1} \\
f^{ij}
&\to& {n_c \; \delta^{ij}\over m} - {\nabla^i \theta_0\over m}\;
\left[ \kappa(a) + {\hbar^2 \; k^2\over4m\;n_c} \right]^{-1}\;
{\nabla^j \theta_0\over m}\,.
\end{eqnarray}
(As desired, this has the net effect of making $f^{\mu\nu}$ a matrix
of numbers, not operators.) The physical wave equation~(\ref{weq-phys})
now becomes a nonlinear dispersion relation
\begin{equation}
f^{00} \;\omega^2 + (f^{0i} +f^{i0}) \;\omega \;k_i +
f^{ij} \;k_i \;k_j = 0.
\end{equation}
After substituting the approximate $D_2$ into this dispersion relation
and rearranging, we see (remember: $k^2 = ||k||^2 = \delta^{ij}
\;k_i \;k_j$)
\begin{equation}
-\omega^2 + 2 \; v_0^i \; \omega k_i
+ {n_c k^2\over m}\left[\kappa(a)+{\hbar^2\over4m n_c} k^2\right] -
(v_0^i \; k_i)^2 = 0.
\end{equation}
That is (with $v_0^i=\frac{1}{m}\nabla_i \theta_0$)
\begin{equation}
\left(\omega - v_0^i \; k_i\right)^2 = {n_c k^2\over
m}\left[\kappa(a)+{\hbar^2\over4m n_c} k^2\right]\,.
\end{equation}
Introducing the speed of sound $\csound$ this takes the form:
\begin{equation} \omega= v_0^i \; k_i \pm \sqrt{\csound^2 k^2+\left({\hbar
\over 2 m}\;k^2\right)^2}.
\label{eq:disprelbec}
\end{equation}
We then see that BEC is a paradigmatic framework where a spacetime geometry emerges at low energies and Lorentz invariance is as an accidental (never exact) symmetry. This symmetry is naturally broken at high energies and appears eminently in modified dispersion relations for the quasi-particles living above the condensate background. 

\section{Modified dispersion relations and their naturalness}
\label{sec:TF}

As mentioned before, not only analogue models but also several QG scenarios played an important role in motivating searcher for departures from Lorentz invariance and  in most of these models, LV enters through modified dispersion relations\index{modified dispersion relations} of the sort \eqref{eq:disprelbec}. These relations can be cast in the general form
\begin{equation}%
E^2=p^2+m^2+f(E,p;\mu;M)\;,%
\label{eq:disprelmod}%
\end{equation}%
where the low energy speed of light $c=1$; $E$ and $p$ are the particle energy and momentum, respectively; $\mu$ is a particle-physics mass-scale (possibly associated with a symmetry breaking/emergence scale) and $M$ denotes the relevant QG scale. Generally, it is assumed that $M$ is of order the Planck mass: $M \sim M_{\rm Pl} \approx 1.22\times 10^{19}\;$GeV, corresponding to a quantum (or emergent) gravity effect.  Note that we assumed a preservation of rotational invariance by QG physics and that only boost invariance is affected by Planck-scale corrections. This does not need to be the case (see however \cite{Jacobson:2005bg} for a discussion about this assumption) and constraints on possible breakdown of rotational invariance have been considered in the literature (especially in the context of the minimal standard model extension). We assume it here only for simplicity and clarity in assessing later the available constraints on the EFT framework.

Of course, once given \eqref{eq:disprelmod} the natural thing to do is to expand the function $f(E,p;\mu;M)$ in powers of the momentum (energy and momentum are basically indistinguishable at high energies, although they are both taken to be smaller than the Planck scale), 
\begin{equation}%
E^2=p^2+m^2+\sum_{i=1}^{\infty} \tilde{\eta}_i p^i\;,%
\label{eq:disprel}%
\end{equation}%
where the lowest order LV terms ($p$, $p^2$, $p^3$, $p^4$) have primarily been considered \cite{Mattingly:2005re}.\footnote{I disregard here the possible appearance of dissipative terms \cite{Parentani:2007uq} in the dispersion relation, as this would correspond to a theory with unitarity loss and to a more radical departure from standard physics than that envisaged in the framework discussed herein (albeit a priori such dissipative scenarios are logically consistent and even plausible within some quantum/emergent gravity frameworks).}

About this last point some comments are in order. In fact, from a EFT point of view the only relevant operators should be the lowest order ones, i.e. those of mass dimension 3,4 corresponding to terms of order $p$ and $p^2$ in the dispersion relation. Situations in which higher order operators ``weight" as much as the lowest order ones are only possible at the cost of a severe, indeed arbitrary, fine tuning of the coefficients $\tilde{\eta}_i$. 

However, we do know by now (see further discussion below) that current observational constraints are tremendous on dimension 3 operators and very severe on dimension 4 ones. This is kind of obvious, given that these operators would end up modifying the dispersion relation of elementary particles at low energies. Dimension 3 operator would dominate at $p\to 0$ while the dimension 4 ones would generically induce a, species dependent, constant shift in the limit speed for elementary particles.
 
Of course one might be content to limit oneself to the study of just these terms but we stress that emergent gravity scenarios, e.g. inspired by analogue gravity models, or QG gravity models, strongly suggest that if the origin of the breakdown of Lorentz invariance is rooted in the UV behaviour of gravitational physics then it should be naturally expected to become evident only at high energies. So one would then predict a hierarchy  of LV coefficients of the sort
\beq \tilde{\eta}_1=\eta_1 \frac{\mu^2}{M},\qquad
\tilde{\eta}_2=\eta_2 \frac{\mu}{M},\qquad \tilde{\eta}_3=\eta_3
\frac{1}{M},\qquad \tilde{\eta}_4=\eta_4 \frac{1}{M^2}.
\label{disprel2} \eeq
In characterizing the strength of a constraint one can then refer to the $\eta_n$ without the tilde, so to compare to what might be expected from Planck-suppressed LV.  In general one can allow the LV parameters $\eta_i$ to depend on the particle type, and indeed it turns out that they {\it   must} sometimes be different but related in certain ways for photon polarization states, and for particle and antiparticle states, if the framework of effective field theory is adopted. 

\subsection{The naturalness problem\index{naturalness problem}}

While the above hierarchy \eqref{disprel2} might seem now a well motivated framework within which asses our investigations, it was soon realized~\cite{Collins:2004bp} that it is still quite unnatural from an EFT point of view. The reason is pretty simple: in EFT radiative corrections will generically allow the percolation of higher dimension Lorentz violation to the lowest dimension terms due to the coupling and self-couplings of particles~\cite{Collins:2004bp}. In EFT loop integrals will be naturally cut-off at the EFT breaking scale, if such scale is as well the Lorentz breaking scale  the two will basically cancel leading to unsuppressed, couplings dependent, contributions to the propagators.  Hence radiative corrections will not allow a dispersion relation with only $p^3$ or $p^4$ Lorentz breaking terms but will automatically induce extra unsuppressed LV terms in $p$ and $p^2$ which will be naturally dominant. 

Several ideas have been advanced in order to justify such a ``naturalness problem" (\see e.g.~\cite{Jacobson:2005bg}), it would be cumbersome to review here all the proposals, but one can clearly see that the most straightforward solution for this problem would consist in breaking the degeneracy between the EFT scale and the Lorentz breaking one.
This can be achieved in two alternative ways. 

\subsubsection{A new symmetry}

Most of the aforementioned proposals implicitly assume that the Lorentz breaking scale is the Planck scale. One then needs the EFT scale (which can be naively identified with what we called previously $\mu$) to be different from the Planck scale and actually sufficiently small so that the lowest order ``induced" coefficients can be suppressed by suitable small rations of the kind $\mu^p/M^q$ where $p,q$ are some positive powers.

A possible solution in this direction can be provided by introducing what is commonly called a ``custodial symmetry" something that forbids lower order operators and, once broken, suppress them by introducing a new scale. The most plausible candidate for this role was soon recognized to be Super Symmetry (SUSY)~\cite{GrootNibbelink:2004za,Bolokhov:2005cj}. SUSY is by definition a symmetry relating fermions to bosons  i.e.~matter with interaction carriers. As a matter of fact, SUSY is intimately related to Lorentz invariance. Indeed, it can be shown that the composition of at least two SUSY transformations induces space-time translations. However, SUSY can still be an exact symmetry even in presence of LV and can actually serve as a custodial symmetry preventing certain operators to appear in LV field theories. 

The effect of SUSY\index{SUSY} on LV is to prevent dimension $\leq 4$, renormalizable LV operators to be present in the Lagrangian.
Moreover, it has been demonstrated \cite{GrootNibbelink:2004za,Bolokhov:2005cj} that the renormalization group equations for Supersymmetric QED plus the addition of dimension 5 LV operators \`a la Myers \& Pospelov \cite{Myers:2003fd} do not generate lower dimensional operators, if SUSY is unbroken. However, this is not the case for our low energy world, of which SUSY is definitely not a symmetry. 

The effect of soft SUSY breaking was also investigated in \cite{GrootNibbelink:2004za,Bolokhov:2005cj}. It was found there that, as expected, when SUSY is broken the renormalizable operators are generated. In particular, dimension $\kappa$ ones arise from the percolation of dimension $\kappa+2$ LV operators\footnote{We consider here only $\kappa = 3,4$, for which these relationships have been demonstrated.}. The effect of SUSY soft-breaking is, however, to introduce a suppression of order $m_{s}^{2}/M_{\rm Pl}$ ($\kappa=3$) or $(m_{s}/M_{\rm Pl})^{2}$ ($\kappa=4$), where $m_{s}\simeq 1$~TeV is the scale of SUSY soft breaking. Although, given present constraints, the theory with $\kappa=3$ needs a lot of fine tuning to be viable, since the SUSY-breaking-induced suppression is not enough powerful to kill linear modifications in the dispersion relation of electrons, if $\kappa = 4$ then the induced dimension 4 terms are suppressed enough, provided $m_{s} < 100$~TeV. Current lower bounds from the Large Hadron Collider are at most around 950 GeV for the most simple models of SUSY~\cite{ATLAS} (the so called ``constrained minimal supersymmetric standard model", CMSSM).

Finally, it is also interesting to note that the analogue model of gravity can be used as a particular implementation of the above mentioned mechanism for avoiding the so called naturalness problem via a custodial symmetry. This was indeed the case of multi-BEC~\cite{Liberati:2005pr,Liberati:2005id}. 

\subsubsection{Gravitational confinement of Lorentz violation}
\label{gravconf}

The alternative to the aforementioned scenario is to turn the problem upside down. One can in fact assume that the Lorentz breaking scale (the $M$ appearing in the above dispersion relations) is not set by the Planck scale while the latter is the EFT breaking scale. If in addition one starts with a theory which has higher order Lorentz violating operators only in the gravitational sector, then one can hope that the gravitational coupling $G_N\sim M^{-2}_{\rm Pl}$ will let them ``percolate" to the matter sector however it will do so introducing factors of the order $(M/M_{\rm Pl})^{2}$ which can become strong suppression factors if $M\ll M_{\rm Pl}$. This is basically the idea at the base of the work presented in \cite{Pospelov:2010mp} which applies it to the special case of Ho\v rava--Lifshitz gravity. There it was shown that indeed a workable low energy limit of the theory can be derived through this mechanism which apparently is fully compatible with extant constraints on Lorentz breaking operators in the matter sector. We think that this new route deserves further attention and should be more deeply explored in the future.

\section{Dynamical frameworks}

Missing a definitive conclusion about the naturalness problem, the study of LV theories has basically proceeded by considering separately extensions of the Standard Model based on naively power counting renormalizable operators or non-renormalizable operators (at some given mass dimension). In what follows we shall succinctly describe these frameworks before to discuss theoretical alternatives. 


\subsection{SME with renormalizable operators}


Most of the research in EFT\index{EFT} with only renormalizable ({\em i.e.}~mass dimension 3 and 4) LV operators has been carried out within the so called (minimal) SME~\cite{KS89}. It consists of the standard model of particle physics plus all Lorentz violating renormalizable operators ({\em i.e.}~of mass dimension $\leq4$) that can be written without changing the field content or violating gauge symmetry. The operators appearing in the SME can be conveniently classified according to their behaviour under CPT. 
Since the most common particles used to cast constraints on LV are photons and electrons, a prominent role is played by LV QED.

If we label by $\pm$ the two photon helicities, we can write the photon dispersion relation as \cite{Kostelecky:2002hh}
\begin{equation}
E = (1+\rho\pm\sigma)|\vec{p}|
\end{equation}
where $\rho$ and $\sigma$ depend on LV parameters appearing in the LV QED Lagrangian, as defined in \cite{Mattingly:2005re}. Note that the dependence of the dispersion relation on the photon helicity is due to the fact that the SME generically also contemplates the possibility of a breakdown of rotational invariance.


We already gave (see section \ref{sec:TF}) motivations for assuming rotation invariance to be preserved, at least in first approximation, in LV contexts.  If we make this assumption, we obtain a major simplification of our framework, because in this case all LV tensors must reduce to suitable products of a time-like vector field, which is usually called $u^{\alpha}$ and, in the preferred frame, is assumed to have components $(1,0,0,0)$. Then, the rotational invariant LV operators are
\begin{equation}
-bu_{\mu}\overline{\psi} \gamma_5 \gamma^{\mu}\psi + \frac {1} {2} i c
u_\mu u_\nu \overline{\psi} \gamma^{\mu}  \stackrel{\leftrightarrow}{D^{\nu}} \psi  + \frac {1} {2} i
d u_\mu u_\nu \overline{\psi} \gamma_5 \gamma^\mu  \stackrel{\leftrightarrow}{D^{\nu}} \psi
\end{equation}
for electrons and
\begin{equation}\label{eq:LVQEDphotonrotinv}
-\frac{1}{4}(k_F){u_\kappa \eta_{\lambda\mu} u_\nu} F^{\kappa\lambda}F^{\mu\nu}
\end{equation}
for photons.  

The high energy ($\Mpl\gg E \gg m$) dispersion relations for QED can be expressed as (see \cite{Mattingly:2005re} and references therein for more details)
\begin{eqnarray} \label{eq:SMErotinvdisp}
E_{\rm el}^2=m_{e}^2+p^2+f^{(1)}_e p+f^{(2)}_ep^2 \quad\mbox{electrons}\\
E_{\gamma}^2=(1+ f^{(2)}_\gamma ){p^2}\quad\mbox{photons}
\end{eqnarray}
where $f^{(1)}_e=-2bs,f^{(2)}_e=-(c-ds)$, and $f^{(2)}_\gamma=k_F/2$ with $s=\pm1$ the helicity state of the electron~\cite{Mattingly:2005re}.
The positron dispersion relation is the same as (\ref{eq:SMErotinvdisp}) replacing $p\rightarrow -p$, this will change only the $f^{(1)}_e$ term.

We notice here that the typical energy at which a new phenomenology should start to appear is quite low. In fact, taking for example $f_{e}^{(2)} \sim O(1)$, one finds that the corresponding extra-term is comparable to the electron mass $m$ precisely at $p \simeq m \simeq 511~\keV$. Even worse, for the linear modification to the dispersion relation, we would have, in the case in which $f^{(1)}_{e} \simeq O(1)$, that $p_{\rm th} \sim m^{2}/\Mpl \sim 10^{-17}~\eV$. (Notice that this energy corresponds by chance to the present upper limit on the photon mass, $m_{\gamma}\lesssim 10^{-18}~\eV$ \cite{pdg}.) 
As said, this implied strong constraints on the parameters and was a further mortivation for exploring the QG preferred possibility of higher order Lorentz violating operators and consequently try to address the naturalness problem.

\subsection{Dimension five operators SME}

An alternative approach within EFT is to study non-renormalizable operators. Nowadays it is widely accepted that the SM could just be an effective field theory and in this sense its renormalizability is seen as a consequence of neglecting some higher order operators which are suppressed by some appropriate mass scale. It is a short deviation from orthodoxy to imagine that such non-renormalizable operators can be generated by quantum gravity effects (and hence be naturally suppressed by the Planck mass) and possibly associated to the violation of some fundamental space-time symmetry like local Lorentz invariance. 

Myers \& Pospelov \cite{Myers:2003fd} found that there are essentially only three operators of dimension five, quadratic in the fields, that can be added to the QED\index{QED} Lagrangian preserving rotation and gauge invariance, but breaking local LI\footnote{Actually these criteria allow the addition of other (CPT even) terms, but these would not lead to modified dispersion relations (they can be thought of as extra, Planck suppressed, interaction terms) \cite{Bolokhov:2007yc}.}.

These extra-terms, which result in a contribution of $O(E/\Mpl)$ to the dispersion relation of the particles, are the following:
\begin{equation}
-\frac{\xi}{2\Mpl}u^mF_{ma}(u\cdot\partial)(u_n\tilde{F}^{na}) + \frac{1}{2\Mpl}u^m\overline{\psi}\gamma_m(\zeta_1+\zeta_2\gamma_5)(u\cdot\partial)^2\psi\:,
\label{eq:LVterms}
\end{equation}
where $\tilde{F}$ is the dual of $F$ and $\xi$, $\zeta_{1,2}$ are dimensionless parameters. All these terms also violate the CPT symmetry.
More recently, this construction has been extended to the whole SM \cite{Bolokhov:2007yc}. 

From (\ref{eq:LVterms}) the dispersion relations of the fields are
modified as follows. For the photon one has
\begin{equation}
\omega_{\pm}^2 = k^2 \pm \frac{\xi}{\Mpl}k^3\:,
\label{eq:disp_rel_phot}
\end{equation}
(the $+$ and $-$ signs denote right and left circular polarisation), while
for the fermion (with the $+$ and $-$ signs now denoting positive and
negative helicity states%
)
\begin{equation}
E_\pm^2 = p^2 + m^2 + \eta_\pm \frac{p^3}{\Mpl}\;,
\label{eq:disp_rel_ferm}
\end{equation}
with $\eta_\pm=2(\zeta_1\pm \zeta_2)$.  For the antifermion, it can be
shown by simple ``hole interpretation" arguments that the same
dispersion relation holds, with $\eta^{af}_\pm = -\eta^f_\mp$ where
$af$ and $f$ superscripts denote respectively anti-fermion and
fermion coefficients~\cite{Jacobson:2005bg,Jacobson:2003bn}.

As we shall see, observations involving very high energies can thus potentially cast $O(1)$ and stronger constraint on the coefficients defined above.  A natural question arises then: what is the theoretically expected value of the LV coefficients in the modified dispersion relations shown above?  

This question is clearly intimately related to the meaning of any constraint procedure. Indeed, let us suppose that, for some reason we do not know, because we do not know the ultimate high energy theory, the dimensionless coefficients $\eta^{(n)}$, that in principle, according to the Dirac criterion, should be of order $O(1)$, are defined up to a dimensionless factor of $m_{e}/\Mpl \sim 10^{-22}$. (This could well be as a result of the integration of high energy degrees of freedom.) Then, any constraint of order larger than $10^{-22}$ would be ineffective, if our aim is learning something about the underlying QG theory. 

This problem could be further exacerbated by renormalization group effects, which could, in principle, strongly suppress the low-energy values of the LV coefficients even if they are $O(1)$ at high energies. Let us, therefore, consider the evolution of the LV parameters first.
Bolokhov \& Pospelov \cite{Bolokhov:2007yc} addressed the problem of calculating the renormalization group equations for QED and the Standard Model extended with dimension-five operators that violate Lorentz Symmetry. 

In the framework defined above, assuming that no extra physics enters between the low energies at which we have modified dispersion relations and the Planck scale at which the full theory is defined, the evolution equations for the LV terms in Eq.~(\ref{eq:LVterms}) that produce modifications in the dispersion relations, can be inferred as
\begin{equation}
\label{eq:RG}
\frac{d\zeta_1}{dt} =  \frac{25}{12}\,\frac{\alpha}{\pi}\,\zeta_1\; , \quad
\frac{d\zeta_2}{dt} =  \frac{25}{12}\,\frac{\alpha}{\pi}\,\zeta_2 - \frac{5}{12}\,\frac{\alpha}{\pi}\,\xi\; , \quad
\frac{d\xi}{dt} =  \frac{1}{12}\,\frac{\alpha}{\pi}\,\zeta_2 - \frac{2}{3}\,\frac{\alpha}{\pi}\,\xi \;,
\end{equation}
where $\alpha = e^2/4\pi \simeq 1/137$ ($\hbar = 1$) is the fine structure constant and $t = \ln(\mu^2/\mu_0^2)$ with $\mu$ and $\mu_0$ two given energy scales. (Note that the above formulae are given to lowest order in powers of the electric charge, which allows one to neglect the running of the fine structure constant.)

These equations show that the running is only logarithmic and therefore low energy constraints are robust: $O(1)$ parameters at the Planck scale are still $O(1)$ at lower energy. Moreover, they also show that $\eta_{+}$ and $\eta_{-}$ cannot, in general, be equal at all scales.

\subsection{Dimension six operators SME}

If CPT ends up being a fundamental symmetry of nature it would forbid all of the above mentioned operators (hence pushing at further high energies the emergence of Lorentz breaking physics). It makes then sense to consider dimension six, CPT even, operators which furthermore do give rise to dispersion relations of the kind appearing in the above mentioned BEC analogue gravity example.

The CPT even dimension 6 LV terms have only recently been computed \cite{Mattingly:2008pw} through the same procedure used by Myers \& Pospelov for dimension 5 LV. The known fermion operators are
\begin{eqnarray}
&\nonumber -\frac{1}{\Mpl}\overline{\psi}(u\cdot D)^{2}(\alpha_{L}^{(5)}P_{L}+\alpha_{R}^{(5)}P_{R})\psi \\
& - \frac{i}{\Mpl^{2}}\overline{\psi}(u\cdot D)^{3}(u\cdot \gamma)(\alpha_{L}^{(6)}P_{L} + \alpha_{R}^{(6)}P_{R}) \psi \\
& \nonumber-\frac{i}{\Mpl^{2}}\overline{\psi} (u\cdot D) \square (u\cdot \gamma) (\tilde{\alpha}_{L}^{(6)}P_{L} + \tilde{\alpha}_{R}^{(6)}P_{R}) \psi\;,
\label{eq:op-dim6-ferm}
\end{eqnarray}
where $P_{R,L}$ are the usual left and right spin projectors $P_{R,L} = (1\pm\gamma^{5})/2$ and where $D$ is the usual QED covariant derivative. All coefficients $\alpha$ are dimensionless because we factorize suitable powers of the Planck mass.

The known photon operator is
\begin{equation}
-\frac{1}{2\Mpl^{2}}\beta_{\gamma}^{(6)}F^{\mu\nu}u_{\mu}u^{\sigma}(u\cdot\partial)F_{\sigma\nu}\;.
\label{eq:op-dim6-phot}
\end{equation}

From these operators, the dispersion relations of electrons and photons can be computed, yielding
\begin{eqnarray}
\nonumber 
E^{2} - p^{2} - m^{2} &=& \frac{m}{\Mpl}(\alpha_{R}^{(5)}+\alpha_{L}^{(5)})E^{2} + \alpha_{R}^{(5)}\alpha_{L}^{(5)}\frac{E^{4}}{\Mpl^{2}} \nonumber \\ 
& & + \frac{\alpha_{R}^{(6)} E^{3}}{\Mpl^{2}}(E+sp) + \frac{\alpha_{L}^{(6)}E^{3}}{\Mpl^{2}}(E-sp) \\
\omega^{2}-k^{2} &=& \beta^{(6)}\frac{k^{4}}{\Mpl^{2}}\;,
\label{eq:disp-rel-dimsix}
\end{eqnarray}
where $m$ is the electron mass and where $s = {\sigma}\cdot\mathbf{p}/|\mathbf{p}|$ is the helicity of the electrons. Also, notice that a term proportional to $E^{2}$ is generated. 

Because the high-energy fermion states are almost exactly chiral, we can further simplify the fermion dispersion relation eq.~(\ref{eq:disp-rel-dimsix}) (we pose $R=+$, $L=-$)
\begin{equation}
E^{2} = p^{2} + m^{2} + f_{\pm}^{(4)} p^{2} + f_{\pm}^{(6)} \frac{p^{4}}{\Mpl^{2}}\;.
\label{eq:disp-rel-ferm-dim6-improved}
\end{equation}
Being suppressed by a factor of order $m/\Mpl$, we will drop in the following the quadratic contribution $f_{\pm}^{(4)}p^{2}$, indeed this can be safely neglected, provided that $E > \sqrt{m\Mpl}$. Let me stress however, that this is exactly an example of a dimension 4 LV term with a natural suppression, which for electron is of order $m_{e}/\Mpl \sim 10^{-22}$. Therefore, any limit larger than $10^{-22}$ placed on this term would not have to be considered as an effective constraint. To date, the best constraint for a rotational invariant electron LV term of dimension 4 is $O(10^{-16})$ \cite{Stecker:2001vb}.

Coming back to equation \eqref{eq:disp-rel-ferm-dim6-improved}, it may seem puzzling that in a CPT invariant theory we distinguish between different fermion helicities. However, although they are CPT invariant, some of the LV terms displayed in Eq.~(\ref{eq:op-dim6-phot}) are odd under P and T. 

CPT invariance allows us to determine a relationship between the LV coefficients of the electrons and those of the positrons. Indeed, to obtain these we must consider that, by CPT, the dispersion relation of the positron is given by (\ref{eq:disp-rel-dimsix}), with the replacements $s \rightarrow -s$ and $p\rightarrow -p$. This implies that the relevant positron coefficients $f^{(6)}_{\rm positron}$ are such that $f^{(6)}_{e^{+}_{\pm}} = f^{(6)}_{e^{-}_{\mp}}$, where $e^{+}_{\pm}$ indicates a positron of positive/negative helicity (and similarly for the $e^{-}_{\pm}$).


\subsection{Other frameworks}
\label{sec:others}

Picking up a well defined dynamical framework is sometimes crucial in discussing the phenomenology of Lorentz violations. In fact, not all the above mentioned ``windows on quantum gravity" can be exploited without adding additional information about the dynamical framework one works with. Although cumulative effects exclusively use the form of the modified dispersion relations, all the other ``windows" depend on the underlying dynamics of interacting particles and on whether or not the standard energy-momentum conservation holds. Thus, to cast most of the constraints on dispersion relations of the form (\ref{eq:disprel}), one needs to adopt a specific theoretical framework justifying  the use of such deformed dispersion relations.  

The previous discussion mainly focuses on considerations based on Lorentz breaking EFTs. This is indeed  a conservative framework within which much can be said (e.g.~reaction rates can still be calculated) and from an analogue gravity point of view it is just the natural frame to work within.  Nonetheless, this is of course not the only dynamical framework within which a Lorentz breaking kinematics can be cast. Because the EFT approach is nothing more than a highly reasonable, but rather arbitrary ``assumption'', it is worth studying and constraining additional models, given that they may evade the majority of the constraints discussed in this review.

\subsubsection{D-brane models}
\label{sec:nonEFT}

We consider here the model presented in \cite{Ellis:1999jf,Ellis:2003sd}, in which modified dispersion relations are found based on the Liouville string approach to quantum space-time \cite{Ellis:1992eh}. Liouville-string models of space-time foam\index{space-time foam} \cite{Ellis:1992eh} motivate corrections to the usual relativistic dispersion relations that are first order in the particle energies and that correspond to a vacuum refractive index $\eta = 1-(E/\Mpl)^{\alpha}$, where $\alpha = 1$. Models with quadratic dependences of the vacuum refractive index on energy: $\alpha = 2$ have also been considered \cite{Burgess:2002tb}.

In particular, the D-particle realization of the Liouville string approach predicts that only gauge bosons such as photons, not charged matter particles such as electrons, might have QG-modified dispersion relations. This difference may be traced to the facts that \cite{Ellis:2003if} excitations which are charged under the gauge group are represented by open strings with their ends attached to the D-brane \cite{Polchinski:1996na}, and that only neutral excitations are allowed to propagate in the bulk space transverse to the brane. Thus, if we consider photons and electrons, in this model the parameter $\eta$ is forced to be null, whereas $\xi$ is free to vary. Even more importantly, the theory is CPT even, implying that vacuum is not birefringent for photons ($\xi_{+} = \xi_{-}$).

\subsubsection{Doubly Special Relativity\index{Doubly Special Relativity}}

Lorentz invariance of physical laws relies on only few assumptions: the principle of relativity, stating the equivalence of physical laws for non-accelerated observers, isotropy (no preferred direction) and homogeneity (no preferred location) of space-time, and a notion of precausality, requiring that the time ordering of co-local events in one reference frame be preserved \cite{ignatowsky,ignatowsky1,ignatowsky2,ignatowsky3,ignatowsky4,Liberati:2001sd,Sonego:2008iu,Baccetti:2011aa}. 

All the realizations of LV we have discussed so far explicitly violate the principle of relativity by introducing a preferred reference frame. This may seem a high price to pay to include QG effects in low energy physics. For this reason, it is worth exploring an alternative possibility that keeps the relativity principle but that relaxes one or more of the above postulates. 

For example, relaxing the space isotropy postulate leads to the so-called Very Special Relativity framework \cite{Cohen:2006ky}, which was later on understood to be described by a Finslerian-type geometry~\cite{Bogoslovsky:2005cs,Bogoslovsky:2005gs,gibbons:081701}. In this example, however, the generators of the new relativity group number fewer than the usual ten associated with Poincar\'e invariance. Specifically, there is an explicit breaking of the $O(3)$ group associated with rotational invariance. 

One may wonder whether there exist alternative relativity groups with the same number of generators as special relativity. Currently, we know of no such generalization in coordinate space. However, it has been suggested that, at least in momentum space, such a generalization is possible, and it was termed  ``doubly" or ``deformed" (to stress the fact that it still has 10 generators) special relativity, DSR.  Even though DSR aims at consistently including dynamics, a complete formulation capable of doing so is still missing, and present attempts face major problems. Thus, at present DSR is only a kinematic theory. Nevertheless, it is attractive because it does not postulate the existence of a preferred frame, but rather deforms the usual concept of Lorentz invariance in the following sense.

Consider the Lorentz algebra of the generators of rotations, $L_i$,
and boosts, $B_i$:%
\begin{equation}%
[L_i,L_j] = \imath\,\epsilon_{ijk}\; L_k\;; \qquad%
[L_i,B_j] = \imath\,\epsilon_{ijk}\; B_k\;; \qquad%
[B_i,B_j] = -\imath\,\epsilon_{ijk}\; L_k%
\label{LB}
\end{equation}%
(Latin indices $i,j,\ldots$ run from 1 to 3) and supplement it with
the following commutators between the Lorentz generators and those of
translations in spacetime (the momentum operators $P_0$ and $P_i$):%
\begin{equation}%
[L_i,P_0]=0\;; \qquad%
[L_i, P_j] = \imath\,\epsilon_{ijk}\; P_k\;;%
\label{LP}%
\end{equation}%
\begin{equation}%
[B_i,P_0] = \imath\; f_1\left(\frac{P}{\kappa}\right) P_i\;;%
\label{BP0}%
\end{equation}%
\begin{equation}%
[B_i,P_j] = \imath\left[ \delta_{ij} \; f_2\left(\frac{P}{\kappa}\right)
P_0 + f_3\left(\frac{P}{\kappa}\right)  \frac{P_i \; P_j}{\kappa}
\right]\;.%
\label{BPj}%
\end{equation}%
where $\kappa$ is some unknown energy scale.
Finally, assume $[P_i,P_j] = 0$. %
%
%
The commutation relations (\ref{BP0})--(\ref{BPj}) are given in terms
of three unspecified, dimensionless structure functions $f_1$, $f_2$,
and $f_3$, and they are sufficiently general to include all known DSR
proposals --- the DSR1~\cite{AmelinoCamelia:2000mn}, DSR2~\cite{Magueijo:2001cr,Magueijo:2002am}, and DSR3~\cite{AmelinoCamelia:2002gv}.  Furthermore, in all the DSRs considered to date, the dimensionless arguments of these functions are specialized to%
\begin{equation}%
f_i\left(\frac{P}{\kappa}\right)  \to f_i\left( \frac{P_0}{\kappa},
\frac{\sum_{i=1}^3 P_i^2}{\kappa^2}\right)\;,%
\end{equation}%
so rotational symmetry is completely unaffected.  For
the $\kappa\to +\infty$ limit to reduce to ordinary special relativity, $f_1$ and $f_2$ must tend to $1$, and $f_3$ must tend to some finite value.%

DSR theory postulates that the Lorentz group still generates space-time symmetries but that it acts in a non-linear way on the fields, such that not only is the speed of light $c$ an invariant quantity, but also that there is a new invariant momentum scale $\kappa$ which is usually taken to be of the order of $\Mpl$. Note that DSR-like features are found in models of non-commutative geometry, in particular in the $\kappa$-Minkowski framework \cite{Lukierski:2004jw,AmelinoCamelia:2008qg}, as well as in non-canonically non commutative field theories \cite{Carmona:2009ra}.

Concerning phenomenology, an important point about DSR in momentum space is that in all three of its formulations (DSR1~\cite{AmelinoCamelia:2000mn}, DSR2~\cite{Magueijo:2001cr,Magueijo:2002am}, and
DSR3~\cite{AmelinoCamelia:2002gv}) the component of the four momentum having deformed commutation with the boost generator can always be rewritten as a non-linear combination of some energy-momentum vector that transforms linearly under the Lorentz group \cite{Judes:2002bw}. For example in the case of DSR2~\cite{Magueijo:2001cr,Magueijo:2002am} one can write
s%
\begin{eqnarray}%
&&E=\frac{-\pi_0}{1-\pi_0/\kappa}\;;%
\label{msE}\\%
&&p_i=\frac{\pi_i}{1-\pi_0/\kappa}\;.%
\label{msp}%
\end{eqnarray}%
It is easy to ensure that while $\pi$ satisfies the usual dispersion
relation $\pi_0^2-\mbox{\boldmath $\pi$}^2=m^2$ (for a particle with
mass $m$), $E$ and $p_i$ satisfy the modified relation%
\begin{equation}%
\left(1-m^2/\kappa^2\right)E^2+2\,\kappa^{-1}\,m^2\,E
-\mbox{\boldmath $p$}^2=m^2\;.%
\label{mod-disp}%
\end{equation}%
Furthermore, a different composition for energy-momentum now holds, given that the composition for the physical DSR momentum $p$ must be derived from the standard energy-momentum conservation of the pseudo-variable $\pi$ and in general implies non-linear terms.
A crucial point is that due to the above structure if a threshold reaction is forbidden in relativistic physics then it is going to be still forbidden by DSR. Hence many constraints that apply to EFT do not apply to DSR.

Despite its conceptual appeal, DSR is riddled with many open problems.
First,  if DSR is formulated as described above --- that is, only in momentum space --- then it is an incomplete theory.  Moreover, because it is always possible to introduce the new variables $\pi_\mu$, on which the
Lorentz group acts in a linear manner, the only way that DSR can avoid
triviality is if there is some physical way of distinguishing the
pseudo-energy $\epsilon\equiv -\pi_0$ from the true-energy $E$, and
the pseudo-momentum $\mbox{\boldmath $\pi$}$ from the true-momentum
$\mbox{\boldmath $p$}$. If not, DSR is no more than a nonlinear choice of coordinates in momentum space.%

In view of the standard relations $E\leftrightarrow \imath\hbar
\partial_t$ and $\mbox{\boldmath $p$}\leftrightarrow -\imath\hbar
\mbox{\boldmath $\nabla$}$ (which are presumably modified in DSR), it is clear that to physically
distinguish the pseudo-energy $\epsilon$ from the true-energy $E$, and
the pseudo-momentum $\mbox{\boldmath $\pi$}$ from the true-momentum
$\mbox{\boldmath $p$}$, one must know how to
relate momenta to position. At a minimum, one needs to develop
a notion of DSR spacetime.%

In this endeavor, there have been two distinct lines of approach, one
presuming commutative spacetime coordinates, the other attempting to
relate the DSR feature in momentum space to a non commutative position
space. In both cases, several authors have pointed out major
problems. In the case of commutative spacetime coordinates, some
analyses have led authors to question the triviality~\cite{Ahluwalia:2002wf} or
internal consistency~\cite{Rembielinski:2002ic,Schutzhold:2003yp,Hossenfelder:2010tm} of DSR.  On the other hand, non-commutative proposals \cite{Lukierski:1993wx} are not yet well understood, although intense research in this direction is under way~\cite{AmelinoCamelia:2007wk}. Finally we cannot omit the recent development of what one could perhaps consider a spin-off of DSR that is Relative Locality, which is based on the idea that the invariant arena for classical physics is a curved momentum space rather than spacetime (the latter being a derived concept)~\cite{AmelinoCamelia:2011bm}.

DSR and Relative Locality are still a subject of active research and debate (see e.g.~\cite{Smolin:2008hd,Rovelli:2008cj,AmelinoCamelia:2011uk,Hossenfelder:2012vk}); nonetheless, they have not yet attained the level of maturity needed to cast robust constraints\footnote{Note however, that some knowledge of DSR phenomenology can be obtained by considering that, as in Special Relativity, any phenomenon that implies the existence of a preferred reference frame is forbidden. Thus, the detection of such a phenomenon would imply the falsification of both special and doubly-special relativity. An example of such a process is the decay of a massless particle.}. For these reasons, in the next sections we focus upon LV EFT and discuss the constraints within this framework.%

\section{Experimental probes of low energy LV: Earth based experiments}

The world as we see it seems ruled by Lorentz invariance to a very high degree. Hence, when seeking tests of Lorentz violations one is confronted with the challenge to find or very high precision experiments able to test Special Relativity or observe effects which might be sensitive to tiny deviations from standard LI. Within the ansatz we lied down in the previous sections it is clear that the first route is practical only when dealing with low energy violations of Lorentz invariance as systematically described by the minimal Standard Model extension (mSME) while astrophysical tests, albeit much less precise, are the choice too for testing LV induced by higher order operators. Let us then briefly review the main experimental tools used so far in order to perform precision tests of Lorentz invariance in laboratory (for more details see e.g.~\cite{Mattingly:2005re,Kostelecky:2008zz,Kostelecky:2008ts}).
 
\subsection{Penning traps}

In a Penning trap a charged particle can be localized for long times using a combination of static magnetic and electric fields. Lorentz violating tests are based on monitoring the particle cyclotron motion in the magnetic field and Larmor precession due to the spin. In fact the relevant frequencies for both these motions are modified in the mSME and Penning traps can be made very sensitive to differences in these frequencies.

\subsection{Clock comparison Experiments}

Clock comparison experiments are generally performed by considering two atomic transition frequencies (which can be considered as two clocks) in the same point in space. The basic idea is that as the clocks move in space, they pick out different components of the Lorentz violating tensors in the mSME. This would yield a sidereal drift between the two clocks. Measuring the difference between the frequencies over long periods, allows to cast very high precision limits on the parameters in the mSME (generally for for protons and neutrons.) 

\subsection{Cavity Experiments}

In cavity experiments one casts constraints on the variation of the cavity resonance frequency as its orientation changes in space. While this is intrinsically similar to clock comparison experiments, these kind of experiments allows to cast constraints also on the electromagnetic sector of the mSME.

\subsection{Spin polarized torsion balance}

The electron sector of the mSME can be effective constrained via spin-torsion balances.
An example is an octagonal pattern of magnets which is constructed so to have an overall spin polarization in the octagonÕs plane.
Four of these octagons are suspended from a torsion fiber in a vacuum chamber. This arrangement of the magnets give an estimated net spin polarization equivalent to $\approx10^{23}$ aligned electron spins. The whole apparatus is mounted on a turntable. As the turntable moves Lorentz violation in the mSME produces an interaction potential for non-relativistic electrons which induces a torque on the torsion balance. The torsion fiber is then twisted by an amount related to the relevant LV coefficients.

\subsection{Neutral mesons}

In the mSME one expects an orientation dependent change in the mass difference e.g. of neutral Kaons.
By looking for sidereal variations or other orientation effects one can derive bounds on each component of the relevant LV coefficients.

\section{Observational probes of high energy LV: astrophysical QED reactions}


Let us begin with a brief review of the most common types of reaction exploited in order to give constraints on the QED sector.

For definiteness, we refer to the following modified dispersion relations:
\begin{eqnarray}
\label{eq:mdr}
E^{2}_{\gamma} &=& k^{2} + \xi_{\pm}^{(n)}\frac{k^{n}}{\Mpl^{n-2}}\qquad \mbox{Photon}\\
E^{2}_{el} &=& m_{e}^{2} + p^{2} + \eta_{\pm}^{(n)}\frac{p^{n}}{\Mpl^{n-2}}\qquad\mbox{Electron-Positron}\;,
\end{eqnarray}
where, in the EFT case, we have $\xi^{(n)} \equiv \xi_{+}^{(n)} = (-)^{n}\xi_{-}^{(n)}$ and $\eta^{(n)} \equiv \eta_{+}^{(n)} = (-)^{n}\eta_{-}^{(n)}$.

\subsection{Photon time of flight\index{Photon time of flight}}
\label{subsec:tof}

Although photon time-of-flight constraints currently provide limits several orders of magnitude weaker than the best ones, they have been widely adopted in the astrophysical community. Furthermore they were the first to be proposed in the seminal paper \cite{AmelinoCamelia:1997gz}.  More importantly, given their purely kinematical nature, they may be applied to a broad class of frameworks beyond EFT with LV. For this reason, we provide a general description of time-of-flight effects, elaborating on their application to the case of EFT below.

In general, a photon dispersion relation in the form of (\ref{eq:mdr}) implies that photons of different colors (wave vectors $k_1$ and $k_2$) travel at slightly different speeds.  Let us first assume that there are no birefringent effects, so that $\xi_{+}^{(n)} = \xi_{-}^{(n)}$.
Then, upon propagation on a cosmological distance $d$, the effect of energy dependence of the photon group velocity produces a time delay
\begin{equation}
 \Delta t^{(n)} = \frac{n-1}{2}\, \frac{k_2^{n-2}-k_1^{n-2}}{\Mpl^{n-2}}\,\xi^{(n)}\, d\;,
\label{eq:tof-naive}
\end{equation}
which clearly increases with $d$ and with the energy difference as long as $n>2$. The largest systematic error affecting this method is the uncertainty about whether photons of different energy are produced simultaneously in the source.

So far, the most robust constraints on $\xi^{(3)}$, derived from time of flight differences, have been obtained within the $D-$brane model (discussed in section \ref{sec:nonEFT}) from a statistical analysis applied to the arrival times of sharp features in the intensity at different energies from a large sample of GRBs with known redshifts~\cite{Ellis:2005wr}, leading to limits $\xi^{(3)}\leq O(10^3)$.
A recent example illustrating the importance of systematic uncertainties can be found in \cite{Albert:2007qk}, where the strongest limit $\xi^{(3)} < 47$ is found by looking at a very strong flare in the TeV band of the AGN Markarian 501. 

One way to alleviate systematic uncertainties --- available only in the context of birefringent theories, such as the one with $n=3$ in EFT --- would be to measure the velocity difference between the two polarization states at a single energy, corresponding to
\begin{equation}
 \Delta t = 2|\xi^{(3)}|k\, d/\Mpl\;.
\end{equation}
This bound would require that both polarizations be observed and that no spurious helicity-dependent mechanism (such as, for example, propagation through a birefringent medium) affects the relative propagation of the two polarization states.

Let us stress that Eq.~(\ref{eq:tof-naive}) is no longer valid in birefringent theories. In fact, photon beams generally are not circularly polarized; thus, they are a superposition of fast and slow modes. Therefore, the net effect of this superposition may partially or completely erase the time-delay effect. To compute this effect on a generic photon beam in a birefringent theory, let us describe a beam of light by means of the associated electric field, and let us assume that this beam has been generated with a Gaussian width
\begin{equation}
\vec{E} = A\, \left(e^{i(\Omega_{0}t-k^{+}(\Omega_{0})z)}\,e^{-(z-v_{g}^{+}t)^{2}\delta\Omega_{0}^{2}}\hat{e}_{+} + e^{i(\Omega_{0}t-k^{-}(\Omega_{0})z)}\,e^{-(z-v_{g}^{-}t)^{2}\delta\Omega_{0}^{2}}\hat{e}_{-}       \right)\;,
\end{equation}
where $\Omega_{0}$ is the wave frequency, $\delta\Omega_{0}$ is the gaussian width of the wave, $k^{\pm}(\Omega_{0})$ is the ``momentum'' corresponding to the given frequency according to (\ref{eq:mdr}) and $\hat{e}_{\pm}\equiv (\hat{e}_{1}\pm i\hat{e}_{2})/\sqrt{2}$ are the helicity eigenstates. Note that by complex conjugation $\hat{e}_{+}^{*} = \hat{e}_{-}$. Also, note that $k^{\pm}(\omega) = \omega \mp \xi \omega^{2}/\Mpl$. Thus,
\begin{equation}
\vec{E} = A\, e^{i\Omega_{0}(t-z)}\left(e^{i\xi\Omega_{0}^{2}/\Mpl z}\,e^{-(z-v_{g}^{+}t)^{2}\delta\Omega_{0}^{2}}\hat{e}_{+} + e^{-i\xi\Omega_{0}^{2}/\Mpl z}\,e^{-(z-v_{g}^{-}t)^{2}\delta\Omega_{0}^{2}}\hat{e}_{-}       \right)\;.
\end{equation}
The intensity of the wave beam can be computed as
\begin{eqnarray}
\nonumber \vec{E}\cdot\vec{E}^{*} &=& |A|^{2}\left( e^{2i\xi\Omega_{0}^{2}/\Mpl z} + e^{-2i\xi\Omega_{0}^{2}/\Mpl z}  \right) e^{-\delta\Omega_{0}^{2}\left( (z-v_{g}^{+}t)^{2} + (z-v_{g}^{-}t)^{2}\right)}\\
&=& 2|A|^{2}e^{-2\delta\Omega_{0}^{2}(z-t)^{2}}\cos\left( 2\xi\frac{\Omega_{0}}{\Mpl}\Omega_{0}z\right)e^{ - 2\xi^{2}\frac{\Omega_{0}^{2}}{M^{2}}(\delta\Omega_{0}t)^{2}}\;.
\end{eqnarray}
This shows that there is an effect even on a linearly-polarised beam. The effect is a modulation of the wave intensity that depends quadratically on the energy and linearly on the distance of propagation. In addition, for a gaussian wave packet, there is a shift of the packet centre, that is controlled by the square of $\xi^{(3)}/\Mpl$ and hence is strongly suppressed with respect to the cosinusoidal modulation.

\subsection{Vacuum Birefringence\index{Vacuum Birefringence}}
\label{sec:birefringence}

The fact that electromagnetic waves with opposite ``helicities'' have slightly different group velocities, in EFT LV with $n=3$, implies that the polarisation vector of a linearly polarised plane wave with energy $k$ rotates, during the wave propagation over a distance $d$, through the angle \cite{Jacobson:2005bg}
\footnote{Note that for an object located at cosmological distance (let $z$ be its redshift), the distance $d$ becomes
\begin{equation}
d(z) = \frac{1}{H_{0}}\int^{z}_0 \frac{1+z'}{\sqrt{\Omega_{\Lambda} + \Omega_{m}(1+z')^{3}}}\,dz'\;,
\end{equation}
where $d(z)$ is not exactly the distance of the object as it includes a $(1+z)^{2}$ factor in the integrand to take into account the redshift acting on the photon energies.}
\begin{equation} 
\theta(d) = \frac{\omega_{+}(k)-\omega_{-}(k)}{2}d \simeq \xi^{(3)}\frac{k^2 d}{2\,M_{\rm Pl}}\;.
\label{eq:theta}
\end{equation} 

Observations of polarized light from a distant source can then lead to a constraint on $|\xi^{(3)}|$ that, depending on the amount of available information --- both on the observational and on the theoretical (i.e.~astrophysical source modeling) side --- can be cast in two different ways \cite{Maccione:2008tq}:
\begin{enumerate}
\item
Because detectors have a finite energy bandwidth, Eq.~(\ref{eq:theta}) is never probed in real situations. Rather, if some net amount of polarization is measured in the band $k_{1} < E < k_{2}$, an order-of-magnitude constraint arises from the fact that if the angle of polarization rotation (\ref{eq:theta}) differed by more than $\pi/2$ over this band,
the detected polarization would fluctuate sufficiently for the net signal polarization to be suppressed \cite{Gleiser:2001rm, Jacobson:2003bn}.  
From (\ref{eq:theta}), this constraint is
\begin{equation} 
\xi^{(3)}\lesssim\frac{\pi\,M_{\rm Pl}}{(k_2^2-k_1^2)d(z)}\;,
\label{eq:decrease_pol}
\end{equation} 
%
This constraint requires that any intrinsic polarization (at source) not be
completely washed out during signal propagation. It thus relies on the
mere detection of a polarized signal; there is no need to consider the observed
polarization degree.
A more refined limit can be obtained by calculating the maximum
observable polarization degree, given the maximum intrinsic value \cite{McMaster}:
\begin{equation} 
\Pi(\xi) = \Pi(0) \sqrt{\langle\cos(2\theta)\rangle_{\mathcal{P}}^{2}
+\langle\sin(2\theta)\rangle_{\mathcal{P}}^{2}},
\label{eq:pol}
\end{equation} 
where $\Pi(0)$ is the maximum intrinsic degree of polarization,
$\theta$ is defined in Eq.~(\ref{eq:theta}) and the average is
weighted over the source spectrum and instrumental efficiency,
represented by the normalized weight function
$\mathcal{P}(k)$~\cite{Gleiser:2001rm}.  
Conservatively, one can set $\Pi(0)=100\%$, but a lower value 
may be justified on the basis of source modeling.
Using \eqref{eq:pol}, one can then 
cast a constraint by 
requiring $\Pi(\xi)$ to exceed the observed value. 

\item
Suppose 
that
polarized light 
measured in a certain energy band 
has
a position angle $\theta_{\rm obs}$ with respect to a fixed
direction. At fixed energy, the polarization vector rotates by the
angle (\ref{eq:theta}) \footnote{Faraday rotation is negligible at
high energies.}; if the position angle is measured by averaging over a
certain energy range, the final net rotation 
$\left<\Delta\theta\right>$
is given by the
superposition of the polarization vectors of all the photons in that
range:
%
%
\begin{equation}
\tan (2\left\langle\Delta\theta\right\rangle) = \frac{
\left\langle\sin(2\theta)\right\rangle_{\mathcal{P}}}{\left\langle
\cos(2\theta)\right\rangle_{\mathcal{P}}}\;,
\label{eq:caseB}
\end{equation}
where 
$\theta$ is given by (\ref{eq:theta}).
If the position angle at emission $\theta_{\rm i}$ in the same energy band 
is known from a model of the emitting source, a constraint can be set by imposing
\begin{equation}
\tan(2\left\langle\Delta\theta\right\rangle) < \tan(2\theta_{\rm obs}-2\theta_{\rm i})\;.
\label{eq:constraint-caseB}
\end{equation}
%
Although this limit is tighter than those based on eqs.~(\ref{eq:decrease_pol}) and (\ref{eq:pol}), it clearly hinges on assumptions about the 
nature of the source, 
which may introduce significant uncertainties.
\end{enumerate}

In conclusion the fact that polarised photon beams are indeed observed from distant objects imposes strong constraints on LV in the photon sector (i.e.~on $\xi^{(3)}$), as we shall see later on.

\subsection{Threshold reactions\index{Threshold reactions}}
\label{sec:thresholds}

An interesting phenomenology of threshold reactions is introduced by LV in EFT; also, threshold theorems can be generalized \cite{Mattingly:2002ba}. Sticking to the present case of rotational invariance and monotonic dispersion relations (see \cite{Baccetti:2011us} for a generalization to more complex situations), the main conclusions of the investigation into threshold reactions are that \cite{Jacobson:2002hd}
\begin{itemize}
\item Threshold configurations still corresponds to head-on incoming particles and parallel outgoing ones
\item The threshold energy of existing threshold reactions can shift, and upper thresholds (i.e.~maximal incoming momenta at which the reaction can happen in any configuration) can appear
\item Pair production can occur with unequal outgoing momenta
\item New, normally forbidden reactions can be viable
\end{itemize}

LV corrections are surprisingly important in threshold reactions because the LV term (which as a first approximation can be considered as an additional mass term) should be compared not to the momentum of the involved particles, but rather to the (invariant) mass of the particles produced in the final state. Thus, an estimate for the threshold energy is
\begin{equation}
p_{\rm th} \simeq \left(\frac{m^{2}\Mpl^{n-2}}{\eta^{(n)}}\right)^{1/n}\;,
\label{eq:threshold-general}
\end{equation}
where $m$ is the typical mass of particles involved in the reaction. Interesting values for $p_{\rm th}$ are discussed, e.g., in \cite{Jacobson:2002hd} and given in Tab.~\ref{tab:thresholds}. 
\begin{table}[htbp]
\caption{Values of $p_{\rm th}$, according to eq.~(\ref{eq:threshold-general}), for different particles involved in the reaction: neutrinos, electrons and proton. Here we assume $\eta^{(n)} \simeq 1$.}
\begin{center}
\begin{tabular}{|c|c|c|c|}
\hline
& $m_{\nu}\simeq 0.1~\eV$ & $m_{e}\simeq 0.5~\MeV$ & $m_{p} \simeq 1~\GeV$ \\
\hline
$n=2$ & 0.1 eV & 0.5 MeV & 1 GeV\\
\hline 
$n=3$ & 500 MeV & 14 TeV & 2 PeV\\
\hline 
$n=4$ & 33 TeV & 74 PeV & 3 EeV\\
\hline
\end{tabular}
\end{center}
\label{tab:thresholds}
\end{table}%
Reactions involving neutrinos are the best candidate for observation of LV effects, whereas electrons and positrons can provide results for $n=3$ theories but can hardly be accelerated by astrophysical objects up to the required energy for $n=4$. In this case reactions of protons can be very effective, because cosmic-rays can have energies well above 3 EeV.
Let us now briefly review the main reaction used so far in order to casts constraints.

\subsubsection{LV-allowed threshold reactions: $\gamma$-decay\index{$\gamma$-decay}} The decay of a photon into an electron/positron pair is made possible by LV because energy-momentum conservation may now allow reactions described by the basic QED vertex. This process has a threshold that, if $\xi \simeq 0$ and $n=3$, is set by the condition \cite{Jacobson:2005bg}
\begin{equation}
k_{th} = (6\sqrt{3}m_e^2M/|\eta_\pm^{(3)}|)^{1/3}\, . 
\end{equation}
Noticeably, as already mentioned above, the electron-positron pair can now be created with slightly different outgoing momenta (asymmetric pair production).
Furthermore, the decay rate is extremely fast above threshold \cite{Jacobson:2005bg} and is of the order of $(10~{\rm ns})^{-1}$ ($n=3$) or $(10^{-6}~{\rm ns})^{-1}$ ($n=4$).

\subsubsection{LV-allowed threshold reactions: Vacuum \v{C}erenkov\index{Vacuum \v{C}erenkov} and Helicity Decay\index{Helicity Decay}} In the presence of LV, the process of Vacuum \v{C}erenkov (VC) radiation $e^{\pm}\rightarrow e^{\pm}\gamma$ can occur. If we set $\xi \simeq 0$ and $n=3$, the threshold energy is given by
\begin{equation}
 p_{\rm VC} = (m_e^2M/2\eta^{(3)})^{1/3} \simeq 11~\TeV~\eta^{-1/3}\;.
\label{eq:VC_th}
\end{equation}
Just above threshold this process is extremely efficient, with a time scale of order $\tau_{\rm VC} \sim 10^{-9}~\s$ \cite{Jacobson:2005bg}. 

A slightly different version of this process is the Helicity Decay (HD, $e_{\mp}\rightarrow e_{\pm}\gamma$). If $\eta_{+} \neq \eta_{-}$, an electron/positron can flip its helicity by emitting a suitably polarized photon. This reaction does not have a real threshold, but rather an effective one \cite{Jacobson:2005bg} --- $ p_{\rm HD} = (m_e^2M/\Delta\eta)^{1/3}$, where $\Delta\eta = |\eta_+^{(3)}-\eta_-^{(3)}|$ --- at which the decay lifetime $\tau_{HD}$ is minimized. For $\Delta\eta\approx O(1)$ this effective threshold is around 10 TeV.  
Note that below threshold $\tau_{\rm HD} > \Delta\eta^{-3} (p/10~\TeV)^{-8}\, 10^{-9}\s$, while above threshold $\tau_{\rm HD}$ becomes independent of $\Delta\eta$~\cite{Jacobson:2005bg}. 

Apart from the above mentioned examples of reactions normally forbidden and now allowed by LV dispersion relations, one can also look for modifications of normally allowed threshold reactions especially relevant in high energy astrophysics.

\subsubsection{LV-allowed threshold reactions: photon splitting and lepton pair production}

It is rather obvious that once photon decay and vacuum \v{C}erenkov are allowed also the related relations in which respectively the our going lepton pair is replaced by two or more photons, $\gamma \rightarrow 2 \gamma$ and $\gamma \rightarrow 3 \gamma$, etc.\ , or the outgoing photons is replaced by an electron-positron pair, $e^- \rightarrow e^-e^-e^+$, are also allowed.

\paragraph{Photon splitting\index{Photon Splitting}}
This is forbidden for $\xi^{(n)}<0$ while it is always allowed if  $\xi^{(n)} > 0$ \cite{Jacobson:2002hd}. When allowed, the relevance of this process is simply related to its rate. The most relevant cases are $\gamma\rightarrow \gamma\gamma$ and $\gamma\rightarrow3\gamma$, because processes with more photons in the final state are suppressed by more powers of the fine structure constant. 

The $\gamma\rightarrow\gamma\gamma$ process is forbidden in QED because of  kinematics and C-parity conservation. In LV EFT neither condition holds. However, we can argue that this process is suppressed by an additional power of the Planck mass, with respect to $\gamma\rightarrow 3 \gamma$. In fact, in LI QED the matrix element is zero due to the exact cancellation of fermionic and anti-fermionic loops. In LV EFT this cancellation is not exact and the matrix element is expected to be proportional to at least $(\xi E/\Mpl)^{p}$, $p>0$, as it is induced by LV and must vanish in the limit $\Mpl \rightarrow \infty$. 

Therefore we have to deal only with $\gamma\rightarrow 3 \gamma$. This process has been studied in \cite{Jacobson:2002hd,Gelmini:2005gy}. In particular, in \cite{Gelmini:2005gy} it was found that, if the ``effective photon mass'' $m_{\gamma}^{2} \equiv \xi E_{\gamma}^{n}/\Mpl^{n-2} \ll m_{e}^{2}$, then the splitting lifetime of a photon is approximately $\tau^{n=3}\simeq 0.025\,\xi^{-5} f^{-1}\left(50~\TeV/E_{\gamma}\right)^{14}~\s$, where $f$ is a phase space factor of order 1. 
This rate was rather higher than the one obtained via dimensional analysis in \cite{Jacobson:2002hd} because, due to integration of loop factors, additional dimensionless contributions proportional to $m_{e}^{8}$ enhance the splitting rate at low energy. 

This analysis, however, does not apply for the most interesting case of ultra high energy photons around $10^{19}$ eV (see below section \ref{Constraints})  given that at these energies $m_{\gamma}^{2} \gg m_{e}^{2}$ if $\xi^{(3)} > 10^{-17}$ and $\xi^{(4)} > 10^{-8}$. Hence the above mentioned loop contributions  are at most logarithmic, as the momentum circulating in the fermionic loop is much larger than $m_{e}$. Moreover, in this regime the splitting rate depends only on $m_{\gamma}$, the only energy scale present in the problem.
One then expects the analysis proposed in \cite{Jacobson:2002hd} to be correct and the splitting time scale to be negligible at $E_{\gamma} \simeq 10^{19}~\eV$.

\paragraph{Lepton pair production}
The process $e^- \rightarrow e^-e^-e^+$ is similar to vacuum
\v{C}erenkov radiation or helicity decay, with the final photon
replaced by an electron-positron pair.   Various combinations of helicities for the different fermions can be considered
individually. If we choose the particularly simple case (and the
only one we shall consider here) where all electrons have the same
helicity and the positron has the opposite helicity, then the threshold energy will depend on
only one LV parameter. In \cite{Jacobson:2002hd} was derived the threshold for this reaction, finding that it is a
factor $\sim 2.5$ times higher than that for soft vacuum \v{C}erenkov radiation. The rate for the reaction is high as well,
hence constraints may be imposed using just the value of the threshold.

\subsubsection{LV-modified threshold reactions: Photon pair-creation\index{Pair-creation}}

A process related to photon decay is photon absorption, $\g\g\rightarrow e^+e^-$. Unlike photon decay,
this is allowed in Lorentz invariant QED and it plays a crucial role in making our universe opaque to gamma rays above tents of TeVs. 

If one of the photons has energy $\omega_0$, the threshold for the reaction occurs in a
head-on collision with the second photon having the momentum
(equivalently energy) $k_{\rm LI}=m^2/\omega_{0}$. For example, if $k_{\rm LI}=10$ TeV (the typical energy of inverse Compton generated photons in some active galactic nuclei) the
soft photon threshold $\omega_0$ is approximately 25 meV, corresponding to a wavelength of 50 microns.

In the presence of Lorentz violating dispersion relations the threshold for this process is in general altered, and the process
can even be forbidden. Moreover, as firstly noticed by Klu\'zniak~\cite{Kluzniak:1999qq} and mentioned before, in some cases there is an upper
threshold beyond which the process does not occur. Physically, this means that at sufficiently high momentum the photon does not carry enough energy to create a pair and simultaneously conserve energy and momentum. Note also, that an upper threshold can only be found in regions of the parameter space in which the $\gamma$-decay is forbidden, because if a single photon is able to create a pair, then {\em a fortiori} two interacting photons will do \cite{Jacobson:2002hd}. 

Let us exploit the above mentioned relation $\eta_{\pm}^{e^{-}} = (-)^{n}\eta_{\mp}^{e^{+}}$ between the electron-positron coefficients, and assume that on average the initial state is unpolarized. In this case, using the energy-momentum conservation, the kinematics equation governing pair production is the following \cite{Jacobson:2005bg}
\begin{equation}
\frac{m^{2}}{k^{n}y \left(1-y\right)} =  
  \frac{4\omega_{b}}{k^{n-1}} + \tilde{\xi} - \tilde{\eta} \left( y^{n-1}+(-)^{n}\left(1-y\right)^{n-1}\right)
   \label{eq:ggscat}
\end{equation}
where $\tilde{\xi}\equiv\xi^{(n)}/M^{n-2}$ and $\tilde{\eta}\equiv\eta^{(n)}/M^{n-2}$ are respectively the photon's and electron's LV coefficients divided by powers of $M$, $0 < y < 1$ is the fraction of momentum carried by either the electron or the positron with respect to the momentum $k$ of the incoming high-energy photon and $\omega_{b}$ is the energy of the target photon. In general the  analysis is rather complicated. In particular it is necessary to sort out whether the thresholds are lower or upper ones, and whether they occur with the same or different pair momenta

\subsection{Synchrotron radiation\index{Synchrotron radiation}} 
\label{synchrotron}
Synchrotron emission is strongly affected by LV, however for Planck scale LV and observed energies, it is a relevant ``window" only for dimension four or five LV QED. We shall work out here the details of dimension five QED ($n=3$) for illustrative reasons (see e.g.~\cite{Altschul:2006pv} for the mSME case). 

In both LI and LV cases \cite{Jacobson:2005bg}, most of the radiation from an electron of energy $E$ is emitted at a critical frequency
\begin{equation}
 \omega_c = \frac{3}{2}eB\frac{\gamma^3(E)}{E} 
\label{eq:omega_sync}
\end{equation}
where $\gamma(E) = (1-v^2(E))^{-1/2}$, and $v(E)$ is the electron
group velocity. 

However, in the LV case, and assuming specifically $n=3$, the electron group velocity is given by
\begin{equation}
 v(E)= \frac{\partial E}{\partial p} =\left(1-\frac{m_e^2}{2p^2}+\eta^{(3)}\frac{p}{M}\right)\,.
\end{equation}
Therefore, $v(E)$ can exceed $1$ if $\eta > 0$ or it can be strictly less
than $1$ if $\eta < 0$. 
This introduces a fundamental difference between particles with positive or negative
LV coefficient $\eta$. 

If $\eta$ is negative the group velocity of the electrons is strictly less than the (low energy) speed of light. This implies that, at sufficiently high energy, $\gamma(E)_{-} < E/m_e$, for all $E$. 
As a consequence, the critical frequency $\omega_c^{-}(\gamma, E)$ is always less than a maximal frequency $\omega_c^{\rm max}$ \cite{Jacobson:2005bg}.
Then, if synchrotron emission up to some frequency $\omega_{\rm obs}$ is observed, one can deduce that the LV coefficient for the corresponding leptons cannot be more negative than the value for which $\omega_c^{\rm max}=\omega_{\rm obs}$. Then, if synchrotron emission up to some maximal frequency $\omega_{\rm obs}$ is observed, one can deduce that the LV coefficient for the corresponding leptons cannot be more negative than the value for which $\omega_c^{\rm max}=\omega_{\rm obs}$, leading to the bound~\cite{Jacobson:2005bg}
\begin{equation}
\eta^{(3)}>-\frac{M}{m_e}\left(\frac{0.34\, eB}{m_e\,\omega_{\rm obs}}\right)^{3/2}\;.
\end{equation}

If $\eta$ is instead positive the leptons can be superluminal. One can show that at energies $E_c \gtrsim 8~\TeV /\eta^{1/3}$, $\gamma(E)$ begins to increase fasters than $E/m_e$ and reaches infinity at a finite energy, which corresponds to the threshold for soft VC emission. The critical frequency is thus larger than the LI one and the spectrum shows a characteristic bump due to the enhanced $\omega_c$. 


\section{Current Constraints on the QED sector}
\label{Constraints}
Let us now come to a brief review of the present constraints on LV QED and in other sectors of the standard model. We shall not spell out the technical details here. These can be found in dedicated, recent, reviews such as ~\cite{Liberati:2009pf}.
 
 \subsection{mSME constraints}
It would be cumbersome to summarize here the constraints on the minimal Standard Model extension (dimension there and four operators) as many parameters characterize the full model. A summary can be found in \cite{Kostelecky:2008ts}. One can of course restrict the mSME to the rotational invariant subset. In this case the model basically coincides with the Coleman-Glashow one~\cite{Coleman:1998ti}. In this case the constraints are quite strong, for example on the QED sector one can easily see that the absence of gamma decay up to 50 TeV provides a constraint of order $10^{-16}$ on the difference between the limit speed of photons and electrons~\cite{Stecker:2001vb}. Constraint up to $O(10^{-22})$ can be achieved on other mSME parameters for dimension four LV terms via precision experiments like Penning traps. 
 
\subsection{Constraints on QED with $O(E/M)$ LV}

It is quite remarkable that a single object can nowadays provide the most stringent constraints for LV QED with $O(E/M)$ modified dispersion relations, this object is the Crab Nebula\index{Crab Nebula} (CN). The CN is a source of diffuse radio, optical and X-ray 
radiation associated with a Supernova explosion observed 
in 1054~A.D. Its distance from Earth is approximately $1.9~\kpc$. 
A pulsar, presumably a remnant of the explosion, is located at the centre of
the Nebula. The Nebula emits an extremely broad-band spectrum (21 decades in frequency, see \cite{Maccione:2007yc} for a comprehensive list of relevant observations) that is produced by two major radiation mechanisms. 
The emission from radio to low energy $\gamma$-rays ($E < 1~\GeV$)
is thought to be synchrotron radiation from relativistic electrons, 
whereas inverse Compton (IC) scattering by these
electrons is the favored explanation for the higher energy 
$\gamma$-rays.   From a theoretical point of view, the current understanding of the whole environment is based on the model presented in \cite{Kennel:1984vf}, which accounts for the general features observed in the CN spectrum. 

 Recently, a claim of $|\xi^{(3)}| \lesssim 2 \times 10^{-7}$ was made using UV/optical polarisation measures from GRBs \cite{Fan:2007zb}. However, the strongest constraint to date comes from a local object. In \cite{Maccione:2008tq} the constraint $|\xi^{(3)}| \lesssim 6 \times 10^{-10}$ at 95\% Confidence Level (CL) was obtained by considering the observed polarization of hard-X rays from the CN \cite{integralpol} (see also \cite{Forot:2008ud}).

\subsubsection{Synchrotron constraint}

How the synchrotron emission processes at work in the CN would appear in a ``LV world'' has been studied in \cite{Jacobson:2002ye,Maccione:2007yc}. There the role of LV in modifying the characteristics of the Fermi mechanism (which is thought to be responsible for the formation of the spectrum of energetic electrons in the CN \cite{Kirk:2007tn}) and the contributions of vacuum \v{C}erenkov and helicity decay were investigated  for $n=3$ LV. This procedure requires fixing most of the model parameters using radio to soft X-rays observations, which are basically unaffected by LV.

\begin{figure}[tbp]
 \sidecaption
 \includegraphics[scale = 0.3, angle = 90]{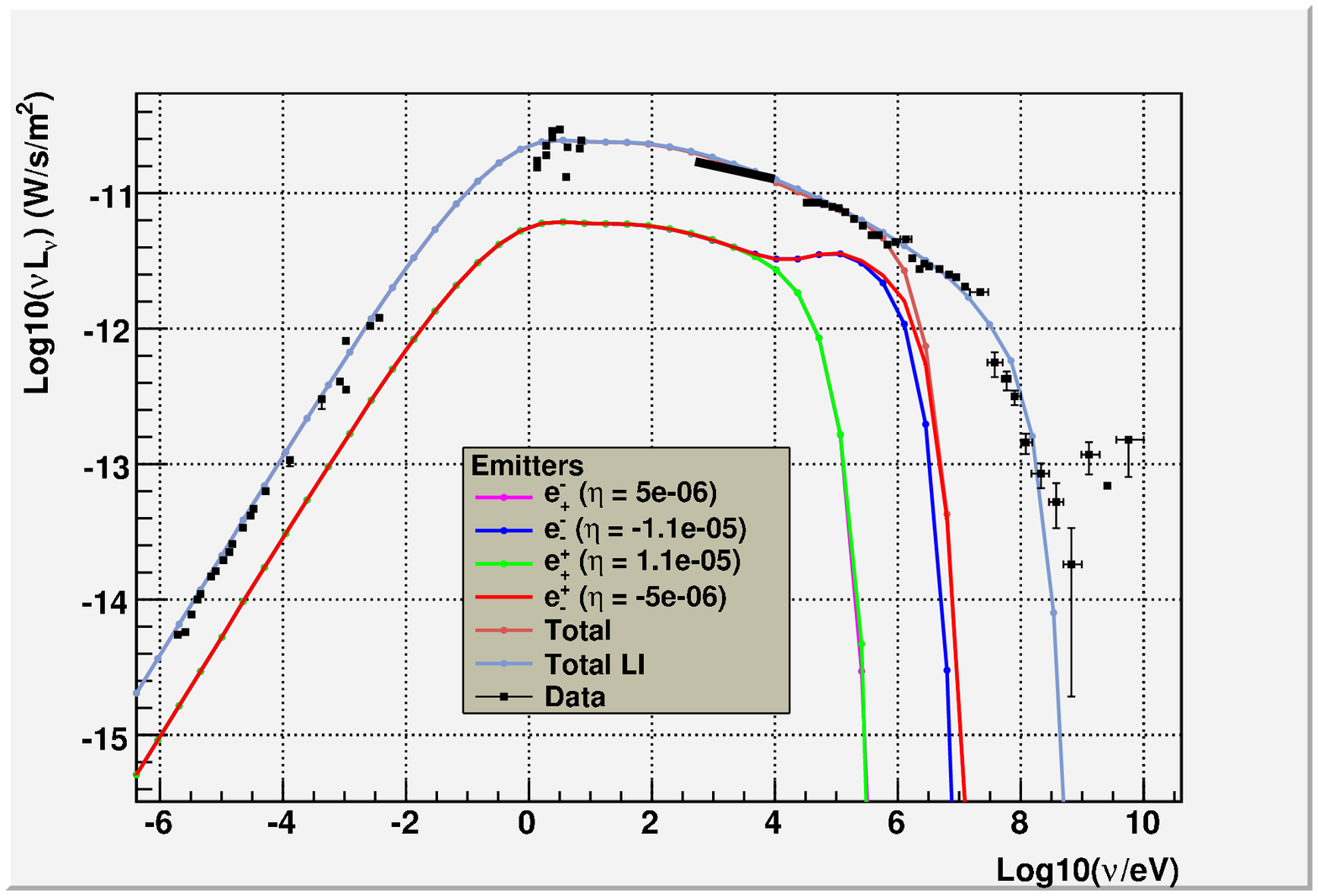}
 \includegraphics[scale = 0.3, angle = 90]{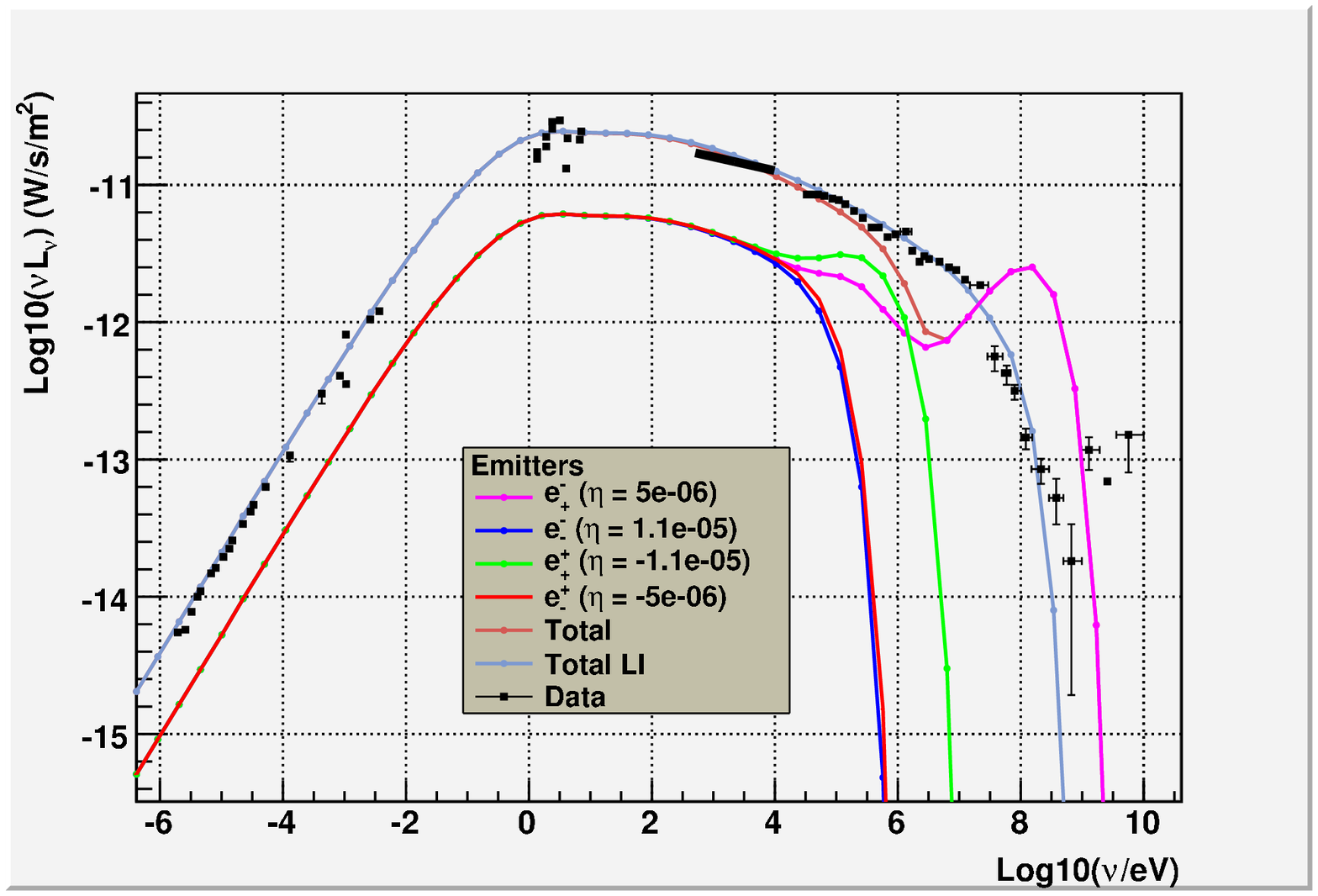}
 \caption{Comparison between observational data, the LI model and a LV
 one with $\eta_+\cdot\eta_- <0$ (left) and $\eta_+\cdot\eta_- >0$
 (right). The values of the LV coefficients, reported in the
 insets, show the salient features
 of the LV modified spectra. The leptons are injected according to the
 best fit values $p=2.4$, $E_c=2.5$ PeV. The individual contribution
 of each lepton population is shown.}
 \label{fig:spectra}
\end{figure}

Given the dispersion relations \eqref{eq:disp_rel_phot} and \eqref{eq:disp_rel_ferm}, clearly only two configurations in the LV parameter space are truly different: $\eta_+\cdot\eta_- >0$ and $\eta_+\cdot\eta_- <0$, where $\eta_+$ is assumed to be positive for definiteness. The configuration wherein both $\eta_\pm$ are negative is the same as the $(\eta_+\cdot\eta_- >0,\,\eta_+>0)$ case, whereas that whose signs are scrambled is equivalent to the case $(\eta_+\cdot\eta_- <0,\,\eta_+>0)$. This is because positron coefficients are related to electron coefficients through $\eta^{af}_\pm = -\eta^{f}_\mp$ \cite{Jacobson:2005bg}. Examples of spectra obtained for the two different cases are shown in Fig.~\ref{fig:spectra}.

A $\chi^2$ analysis has been performed to quantify the agreement between models and data \cite{Maccione:2007yc}.
From this analysis, one can conclude that the LV parameters for the leptons are both constrained, at 95\% CL, to be $|\eta_\pm| < 10^{-5}$, as shown by the red vertical lines in Fig.~\ref{fig:total-n3}. 
Although the best fit model is not the LI one, a careful statistical analysis (performed with present-day data) shows that it is statistically indistinguishable from the LI model at 95\% CL \cite{Maccione:2007yc}. 

\subsubsection{Birefringence constraint}

\begin{figure}[tbp]
\sidecaption[t]
\includegraphics[scale=0.4]{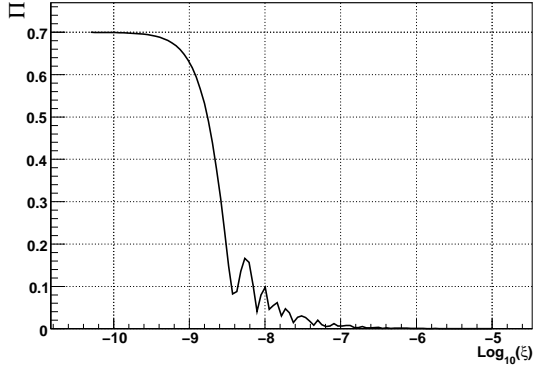} 
\caption{Constraint for the polarization degree. Dependence of $\Pi$
on $\xi$ for the distance of the CN and photons in the 150--300 keV
range, for a constant instrumental sensitivity $\mathcal{P}(k)$.}
\label{fig:caseA}
\end{figure}

In the case of the CN a $(46\pm10)$\%
degree of linear polarization in the $100~\keV - 1~\MeV$ band has 
recently been measured 
by the INTEGRAL mission
\cite{integral,integralpol}.  This 
measurement uses all photons within the SPI instrument energy band. However
the convolution of the instrumental sensitivity to polarization with
the detected number counts as a function of energy, $\mathcal{P}(k)$,
is maximized and approximately constant within a narrower energy band
(150 to 300 keV) and falls steeply outside this range \cite{McGlynn:2007pz}.  For this reason we shall,
conservatively, assume that most polarized photons are concentrated in
this band.
Given $d_{\rm Crab}=1.9~\kpc$, $k_2 = 300~\keV$ and $k_1 = 150~\keV$, 
eq.~(\ref{eq:decrease_pol}) leads to the
order-of-magnitude estimate $|\xi| \lesssim 2\times10^{-9}$.
A more accurate limit follows from 
(\ref{eq:pol}).
In the case of the CN there is
a robust understanding that photons in the range of interest are
produced via the synchrotron proces, for which the maximum degree of
intrinsic linear polarization is about $70\%$ (see e.g.~\cite{Petri:2005ys}).
Figure \ref{fig:caseA} illustrates the dependence of $\Pi$ on $\xi$ (see eq.\eqref{eq:pol})
for the distance of the CN and for $\Pi(0)=70\%$. 
The requirement $\Pi(\xi)>16\%$
(taking account of a $3\sigma$ offset from the best fit value $46\%$) leads to
the constraint (at 99\% CL)
%
\begin{equation}
|\xi| \lesssim 6\times 10^{-9}\;.
\label{eq:constraint-degree}
\end{equation}
It is interesting to notice that X-ray polarization
measurements of the CN already available in 1978
\cite{1978ApJ...220L.117W}, set a constraint $|\xi| \lesssim
5.4\times10^{-6}$, only one order of magnitude less stringent than
that reported in \cite{Fan:2007zb}.

Constraint (\ref{eq:constraint-degree}) can be tightened
by exploiting the current astrophysical understanding of the source.
The CN is a cloud of relativistic particles and fields powered
by a rapidly rotating, strongly magnetized neutron star. Both the {\em
Hubble Space Telescope} and the {\em Chandra} X-ray satellite have
imaged the system, revealing a jet and torus that clearly identify the
neutron star rotation axis \cite{Ng:2003fm}. The projection of this
axis on the sky lies at a position angle of
$124.0^{\circ}\pm0.1^{\circ}$ (measured from North in
anti-clockwise). The neutron star itself emits pulsed radiation at its
rotation frequency of 30 Hz. In the optical band these pulses are
superimposed on a fainter steady component with a linear polarization
degree of ~30\% and direction precisely aligned with that of the
rotation axis \cite{Kanbach:2005kf}.  The direction of polarization
measured by INTEGRAL-SPI in the $\gamma$-rays is $\theta_{\rm obs} =
123^{\circ}\pm11^{\circ}$ ($1\sigma$ error) from the North, thus also
closely aligned with the jet direction and remarkably consistent with
the optical observations.

This compelling (theoretical and observational) evidence allows us to
use eq.~(\ref{eq:constraint-caseB}). Conservatively assuming
$\theta_{\rm i}-\theta_{\rm obs} = 33^{\circ}$
(i.e.~$3\sigma$ from $\theta_{\rm i}$, 99\% CL), this translates into
the limit
\begin{equation}
|\xi^{(3)}| \lesssim 9\times10^{-10}\;,
\label{eq:constraint-serious-crab}
\end{equation}
and $|\xi^{(3)}| \lesssim 6\times10^{-10}$ for a $2\sigma$ deviation (95\%
CL). 

Polarized light from GRBs has also been detected and given their cosmological distribution they could be ideal sources for improving the above mentioned constraints from birefringence. Attempts in this sense were done in the past \cite{Jacobson:2003bn,Mitro} (but later on the relevant observation~\cite{CB} appeared controversial) but so far we do not have sources for which the polarization is detected and the spectral redshift is precisely determined. In \cite{Stecker:2011ps} this problem was circumvented by using indirect methods (the same used to use GRBs as standard candles) for the estimate of the redshift. This leads to a possibly less robust but striking constraints $|\xi^{(3)}| \lesssim 2.4\times10^{-14}$. 

Remakably this constraint was recently further improved by using the INTEGRAL/IBIS observation of the GRB 041219A, for which a luminosity distance of 85 Mpc ($z\approx 0.02$) was derived thanks to the determination of the GRB's host galaxy. In this case a constraint $|\xi^{(3)}| \lesssim 1.1\times10^{-14}$ was derived \cite{Laurent:2011he}.\footnote{The same paper claims also a strong constraint on the parameter $\xi^{(4)}$. Unfortunately, such a claim is based on the erroneous assumption that the EFT order six operators responsible for this term imply opposite signs for opposite helicities of the photon. We have instead seen that the CPT evenness of the relevant dimension six operators imply a helicity independent dispersion relation for the photon (see eq.\eqref{eq:disp-rel-dimsix}).}

\subsubsection{Summary}
\begin{figure}[tbp]
\sidecaption[t]
\includegraphics[scale=0.30]{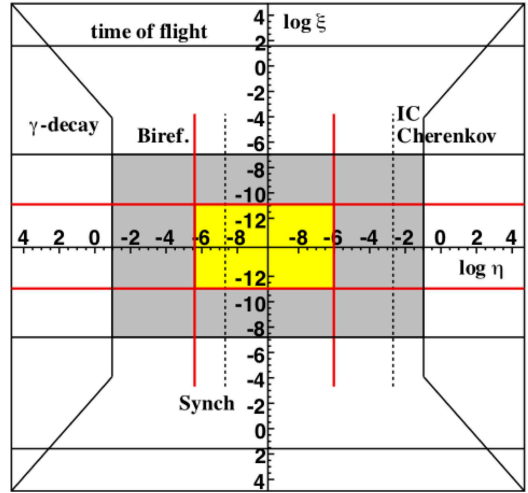} 
\caption{Summary of the constraints on LV QED at order $O(E/M)$. The red lines are related to the constraints derived from the detection of polarized synchrotron radiation from the CN as discussed in the text. For further reference are also shown the constraints that can be derived from the detection of 80 TeV photons from the CN: the solid black lines symmetric w.r.t. the $\xi$ axis are derived from the absence of gamma decay, the dashed vertical line cutting the $\eta$ axis at about $10^{-3}$ refers to the limit on the vacuum \v{C}erenkov effect coming from the inferred 80 TeV inverse Compton electrons. The dashed vertical line on the negative side of the $\eta$ axis is showing the first synchrotron based constraint derived in~\cite{Jacobson:2002ye}.}
\label{fig:total-n3}
\end{figure}

Constraints on LV QED $O(E/M)$ are summarized in Fig.\ref{fig:total-n3} where also the constraints --- coming from the observations of up to 80 TeV gamma rays from the CN \cite{Aharonian:2004gb} (which imply no gamma decay for these photons neither vacuum Cherenkov at least up to 80 TeV for the electrons producing them via inverse Compton scattering) --- are plotted for completeness.

\subsection{Constraints on QED with $O(E/M)^2$ LV}

Looking back at Table \ref{tab:thresholds} it is easy to realize that casting constraints on dimension six LV operators in QED requires accessing energies beyond $10^{16}$ eV. Due to the typical radiative processes characterizing electrons and photons it is extremely hard to directly access these kind of energies. However, the cosmic rays spectrum does extend in this ultra high energy region and it is therefore the main (so far the only) channel for probing these kind of extreme UV LV.

One of the most interesting features related to the physics of Ultra-High-Energy Cosmic Rays\index{Ultra-High-Energy Cosmic Rays} (UHECRs) is the Greisen-Zatsepin-Kuzmin (GZK\index{GZK}) cut off \cite{Greisen:1966jv,1969cora...11...45Z}, a suppression of the high-energy tail of the UHECR spectrum arising from interactions with CMB photons, according to $p\gamma\rightarrow \Delta^{+}\rightarrow p\pi^{0}(n\pi^{+})$. This process has a (LI) threshold energy $E_{\rm th} \simeq 5\times 10^{19}~(\omega_{b}/1.3~\meV)^{-1}~\eV$ ($\omega_{b}$ is the target photon energy). Experimentally, the presence of a suppression of the UHECR flux was claimed only recently \cite{Abbasi:2007sv,Roth:2007in}. Although the cut off could be also due to the finite acceleration power of the UHECR sources, the fact that it occurs at the expected energy favors the GZK explanation. The results presented in \cite{Cronin:2007zz} seemed to further strengthen this hypothesis (but see further discussion below).

Rather surprisingly, significant limits on $\xi$ and $\eta$ can be derived by considering UHE photons generated as secondary products of the GZK reaction\cite{Galaverni:2007tq,Maccione:2008iw}. This can be used to further improve the constraints on dimension 5 LV operators and provide a first robust constraint of QED with dimension 6 CPT even LV operators. 

These UHE photons originate because the GZK process leads to the production of neutral pions that subsequently decay into photon pairs. These photons are mainly absorbed by pair production onto the CMB and radio background. Thus, the fraction of UHE photons in UHECRs is theoretically predicted to be less than 1\% at $10^{19}~\eV$ \cite{Gelmini:2005wu}. Several experiments imposed limits on the presence of photons in the UHECR spectrum. In particular, the photon fraction is less than 2.0\%, 5.1\%, 31\% and 36\% (95\% C.L)~at $E = 10$, 20, 40, 100 EeV  respectively \cite{Aglietta:2007yx,Rubtsov:2006tt}. 

The point is that pair production is strongly affected by LV. In particular, the (lower) threshold energy can be slightly shifted and in general an upper threshold can be introduced \cite{Jacobson:2002hd}. If the upper threshold energy is lower than $10^{20}~\eV$, then UHE photons are no longer attenuated by the CMB and can reach the Earth, constituting a significant fraction of the total UHECR flux and thereby violating experimental limits \cite{Galaverni:2007tq,Maccione:2008iw,Galaverni:2008yj}. 

Moreover, it has been shown \cite{Maccione:2008iw} that the $\gamma$-decay process can also imply a significant constraint. Indeed, if some UHE photon ($E_{\gamma}\simeq 10^{19}~\eV$) is detected by experiments (and the Pierre Auger Observatory, PAO, will be able to do so in few years \cite{Aglietta:2007yx}), then $\gamma$-decay must be forbidden above $10^{19}~\eV$. 

\begin{figure}[tbp]
\sidecaption[t]
 \includegraphics[scale=0.4, angle = 90]{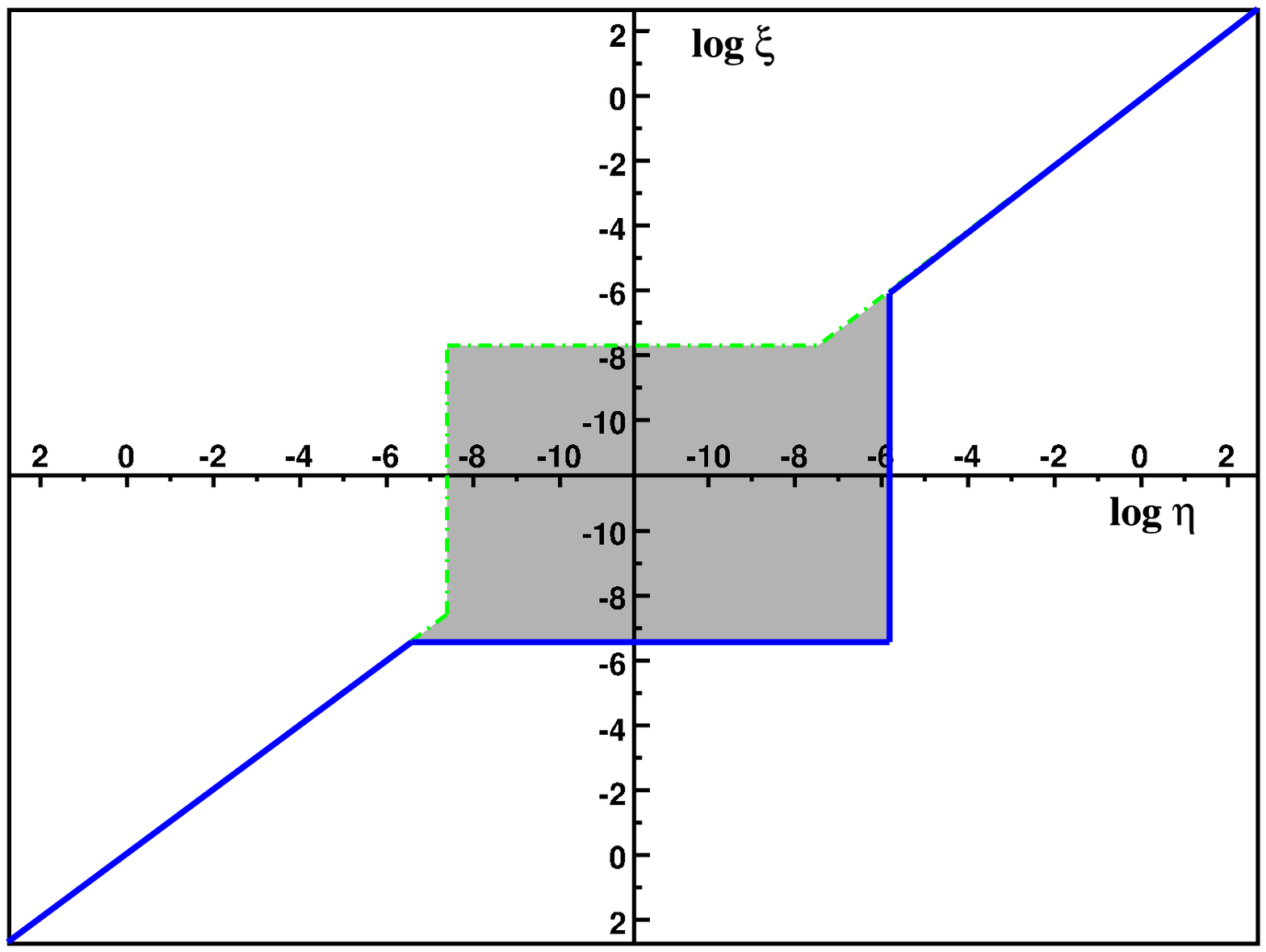}
 \caption{
 LV induced by dimension 6 operators. The LV parameter space is shown. The allowed regions are shaded grey. Green dotted lines represent values of $(\eta, \xi)$ for which the $\gamma$-decay threshold $k_{\gamma-dec} \simeq 10^{19}~\eV$. Solid, blue lines indicate pairs $(\eta,\xi)$ for which the pair production upper threshold $k_{\rm up} \simeq 10^{20}~\eV$.}
 \label{fig:constraints}
\end{figure}

In conclusion we show in Fig.~\ref{fig:constraints} the overall picture of the constraints of QED dimension 6 LV operators, where the green dotted lines do not correspond to real constraints, but to the ones that will be achieved when AUGER will observe, as expected, some UHE photon.

Let us add that the same reasoning can be used to further strength the available constraints in dimension 5 LV QED. In this case the absence of relevant UHE photon flux strengthen by at most two order of magnitude the constraint on the photon coefficient while the eventual detection of the expected flux of UHE photons would constraint the electron positron coefficients down to $|\eta^{(3)}|\lesssim 10^{-16}$ (see \cite{Maccione:2008iw, Liberati:2009pf} for further details) by limiting the gamma decay process (note however, that in this case one cannot exclude that only one photon helicity survives and hence a detailed flux reconstruction would be needed).

\section{Other SM sectors constraints}

While QED constraints are up to date the more straightforward from a theoretical as well observational point of view, it is possible to cast constraints also on other sectors of the SM, most noticeably on the hadronic and neutrino sectors. Let us review them very briefly here.

\subsection{Constraints on the hadronic sector}

Being an ultra high energy threshold process, the aforementioned GZK photopion production is strongly affected by LV. Several authors have studied the constraints implied by the detection of  this effect \cite{Aloisio:2000cm,Alfaro:2002ya,Jacobson:2002hd,Mattingly:2008pw,Scully:2008jp,Stecker:2009hj}. However, a detailed LV study of the GZK feature is hard to perform, because of the many astrophysical uncertainties related to the modeling of the propagation and the interactions of UHECRs.

As a consequence of LV, the mean free path for the GZK reaction is modified. The propagated UHECR spectrum can therefore display features, like bumps at specific energies, suppression at low energy, recovery at energies above the cutoff, such that the observed spectrum cannot be reproduced. Moreover, the emission of Cherenkov $\gamma$-rays and pions in vacuum would lead to sharp suppression of the spectrum above the relevant threshold energy. After a detailed statistical analysis of the agreement between the observed UHECR spectrum and the theoretically predicted one in the presence of LV and assuming pure proton composition, the final constraints implied by UHECR physics are (at
99\% CL) \cite{Maccione:2009ju}
\begin{eqnarray}
\nonumber
-10^{-3} \lesssim &\eta^{(4)}_{p}& \lesssim 10^{-6}\\
-10^{-3} \lesssim &\eta^{(4)}_{\pi}&  \lesssim 10^{-1} \quad (\eta^{(4)}_{p} > 0) \quad\mbox{or}\quad \lesssim 10^{-6} \quad (\eta^{(4)}_{p} < 0)\,.
\label{eq:finalconstraint}
\end{eqnarray}
Of course for dimension five operators much stronger constraints can be achieved by a similar analysis (order $O(10^{-14})$).

\subsection{Constraints on the neutrino sector}

LV can affect the speed of neutrinos with respect to light, influence possible threshold reactions or modified the oscillations between neutrinos flavors. Unfortunately we have a wealth of information only about the latter phenomenon which however constraints only the differences among LV coefficients of different flavors. In this case, the best constraint to date comes from survival of atmospheric muon neutrinos observed by the former IceCube detector AMANDA-II in the energy range 100 GeV to 10 TeV \cite{Kelley:2009zza}, which searched for a generic LV in the neutrino sector~\cite{GonzalezGarcia:2004wg} and achieved $(\Delta c /c)_{ij} \leq 2.8\times10^{-27}$ at 90\% confidence level assuming maximal mixing for some of the combinations $i,j$. Given that IceCube does not distinguish neutrinos from antineutrinos, the same constraint applies to the corresponding antiparticles. The IceCube detector is expected to improve this constraint to $(\Delta c / c)_{ij} \leq 9\times 10^{-28}$ in the next few years \cite{Huelsnitz:2009zz}. The lack of sidereal variations in the atmospheric neutrino flux also yields comparable constraints on some combinations of SME parameters \cite{Abbasi:2010kx}.

For what regards the time of flight constraints we have to date only a single event to rely on, the supernova SN1987a. This was a peculiar event which allowed to detect the almost simultaneous (within a few hours) arrival of electronic antineutrinos and photons. Although only few electronic antineutrinos at MeV energies was detected by the experiments KamiokaII, IMB and Baksan, it was enough to establish a constraint $(\Delta c/c)^{TOF} \lesssim 10^{-8}$ \cite{Stodolsky:1987vd} or $(\Delta c/c)^{TOF} \lesssim 2\times10^{-9}$ \cite{Longo:1987gc} by looking at the difference in arrival time between antineutrinos and optical photons over a baseline distance of $1.5\times10^5$ ly. Further analyses of the time structure of the neutrino signal strengthened this constraint down to $\sim10^{-10}$ \cite{Ellis:2008fc,Sakharov:2009sh}. 

The scarcity of the detected neutrino did not allow the reconstruction of the full energy spectrum and of its time evolution in this sense one should probably consider constraints purely based on the difference in the arrival time with respect to photons more conservative and robust. Unfortunately adopting $\Delta c/c \lesssim 10^{-8}$, the SN constraint implies very weak constraints, $\xi_{\nu}^{(3)}\lesssim 10^{13}$ and $\xi_{\nu}^{(4)}\lesssim 10^{34}$.

Threshold reactions also can be used to cast constraints on the neutrinos sector. In the literature have been considered several processes most prominently the neutrino \v{C}erekov emission $\nu\to\gamma\,\nu$, the neutrino splitting $\nu\to \nu\,\nu \overline{\nu}$ and the neutrino electron/positron pair production $\nu\to\nu \,e^{-}e^{+}$. Let us consider for illustration the latter process. Neglecting possible LV modification in the electron/positron sector (on which we have seen we have already strong constraints) the threshold energy is for arbitrary $n$
\begin{equation}
E_{th, (n)}^{2} = \frac{4m_{e}^{2}}{\delta_{(n)}}\;,
\end{equation}
with $\delta_{(n)} = \xi_{\nu}(E_{th}/M)^{n-2}$.

The rate of this reaction was firstly computed in \cite{Cohen:2011hx} for $n=2$ but can be easily generated to arbitrary $n$~\cite{Maccione:2011fr} (see also \cite{Carmona:2012tp}). The generic energy loss time-scale then reads (dropping purely numerical factors)

\begin{equation}
\tau_{\nu{\rm -pair}} \simeq \frac{ m_Z^4 \cos^4 \theta_w}{g^4E^5} \left( \frac {M} {E} \right)^{3(n-2)}\;,
\end{equation}
where $g$ is the weak coupling and $\theta_w$ is Weinberg's angle.

The observation of upward-going atmospheric neutrinos up to 400 TeV by the experiment IceCube implies that the free path of these particles is at least longer than the Earth radius implies a constraint $\eta^{(3)}_\mu\lesssim 30$. No effective constraint can be optioned for $n=4$ LV, however in this case neutrino splitting (which has the further advantage to be purely dependent on LV on the neutrinos sector) could be used on the ``cosmogenic'' neutrino flux. This is supposedly created via the decay of charged pions produced by the aforementioned GZK effect. The neutrino splitting should modify the spectrum of the ultra high energy neutrinos by suppressing the flux at the highest energies and enhancing it at the lowest ones. In \cite{Mattingly:2009jf} it was shown that future experiments like ARIANNA \cite{Barwick:2006tg} will achieve the required sensitivity to cast a constraint of order $\eta^{(4)}_{\nu} \lesssim 10^{-4}$. Note however, that the rate for neutrino splitting computed in \cite{Mattingly:2009jf} was recently recognized to be  underestimated by a factor $O(E/M)^2$~\cite{Ward:2012fy}. Hence the future constraints here mentioned should be recomputed and one should be able to strengthened them by few orders of magnitude.

\section{Summary and Perspectives}

We can summarize the current status of the constraints for the LV SME in the following table.
\begin{table}[!htb]
\caption{Summary of typical strengths of the available constrains on the SME at different orders.}
\label{tab:1}       
%
%
\begin{center}
\begin{tabular}{p{1.2cm}|p{2.5cm}|p{2.5cm}|p{2.5cm}|p{2.5cm}}
\hline\noalign{\smallskip}
Order & photon & $e^{-}/e^{+}$ & Hadrons & Neutrinos$^a$  \\
\noalign{\smallskip}\svhline\noalign{\smallskip}
n=2 & N.A. & $O(10^{-16})$ & $O(10^{-27})$ & $O(10^{-8})$ \\
n=3 & $O(10^{-14})$ (GRB) & $O(10^{-16})$ (CR) 
& $O(10^{-14})$ (CR) & $O(30)$ \\
n=4 & $O(10^{-8})$ (CR) & $O(10^{-8})$ (CR) & $O(10^{-6})$ (CR)  & $O(10^{-4})^*$ (CR) \\
\noalign{\smallskip}\hline\noalign{\smallskip}
\end{tabular}
\end{center}
GRB=gamma rays burst, CR=cosmic rays\\
$^a$ From neutrino oscillations we have constraints on the difference of LV coefficients of different flavors up to $O(10^{-28})$ on dim 4, $O(10^{-8})$ and expected up to $O(10^{-14})$  on dim 5 (ICE3), expected up to $O(10^{-4})$ on dim 6 op.
$^*$ Expected constraint from future experiments.  
\label{table:sum} 
\end{table}

A special caveat it is due in the case of $n=4$ constraints. As we have seen, they mostly rely (in the QED and Hadronic sector) on the actual detection of the GZK feature of the UHECR spectrum. More specifically, UHECR constraints have relied so far on the hypothesis, not in contrast with any previous experimental evidence, that protons constituted the majority of UHECRs above $10^{19}~\eV$. Recent PAO \cite{Abraham:2010yv} and Yakutsk \cite{Glushkov:2007gd} observations, however, showed hints of an increase of the average mass composition with rising energies up to $E \approx 10^{19.6}~\eV$, although still with large uncertainties mainly due to the proton-air cross-section at ultra high energies. Hence, experimental data suggests that heavy nuclei can possibly account for a substantial fraction of UHECR arriving on Earth. 

Furthermore the evidence for correlations between UEHCR events and their potential extragalactic sources \cite{Cronin:2007zz}--- such as active galactic nuclei (mainly Blasars) --- has not improved with increasing statistics. This might be interpreted as a further hint that a relevant part of the flux at very high enrages should be accounted for by heavy ions (mainly iron) which are much more deviated by the extra and inter galactic magnetic fields due to their larger charge with respect to protons (an effect partially compensated by their shorter mean free path at very high energies). 

If consequently one conservatively decides to momentarily suspend his/her judgment about the evidence for a GZK feature, then he/she would lose the constraints at $n=4$ on the QED sector\footnote{This is a somewhat harsh statement given that it was shown in \cite{Hooper:2010ze} that a substantial (albeit reduced) high energy gamma ray flux is still expected also in the case of mixed composition, so that in principle the previously discussed line of reasoning based on the absence of upper threshold for UHE gamma rays might still work.} as well as very much weaken the constraints on the hadronic one. 

Assuming that current hints for a heavy composition at energies $E \sim 10^{19.6}~\eV$ \cite{Abraham:2010yv} may be confirmed in the future, that some UHECR is observed up to $E \sim 10^{20}~\eV$ \cite{Abraham:2010mj}, and that the energy and momentum of the nucleus are the sum of energies and momenta of its constituents (so that the parameter in the modified dispersion relation of the nuclei is the same of the elementary nucleons, specifically $\eta_p$) one could place a first constraint on the absence of spontaneous decay for nuclei which could not spontaneously decay without LV.\footnote{UHE nuclei suffer mainly from photo-disintegration losses as they propagate in the intergalactic medium. Because photo-disintegration is indeed a threshold process, it can be strongly affected by LV. According to \cite{Saveliev:2011vw}, and in the same way as for the proton case, the mean free paths of UHE nuclei are modified by LV in such a way that the final UHECR spectra after propagation can show distinctive LV features. However, a quantitative evaluation of the propagated spectra has not been performed yet.}

It will place a limit on $\eta_{p}<0$, because in this case the energy of the emitted nucleon is lowered with respect to the LI case until it ``compensates'' the binding energy of the nucleons in the initial nucleus in the energy-momentum conservation. An upper limit for $\eta_{p}>0$ can instead be obtained from the absence of vacuum Cherenkov emission.  If UHECR are mainly iron at the highest energies the constraint is given by $\eta_{p} \lesssim 2\times10^{2}$ for nuclei observed at $10^{19.6}~\eV$ (and $\eta_{p} \lesssim 4$ for $10^{20}~\eV$), while for helium it is $\eta_{p} \lesssim 4\times10^{-3}$ \cite{Saveliev:2011vw}.

So, in conclusion, we can see that the while much has been done still plenty is to be explored. In particular, all of our constraints on $O(E/M)^2$ LV EFT (the most interesting order from a theoretical point of view) are based on the GZK effect (more or less directly) whose detection is still uncertain. It would be nice to be able to cast comparable constrains using more reliable observations, but at the moment it is unclear what reaction could play this role. Similarly, new ideas like the one of gravitational confinement \cite{Pospelov:2010mp} presented in section \ref{gravconf}, seems to call for much deeper investigation of LV phenomenology in the purely gravitational sector.  

We have gone along way into exploring the possible phenomenology of Lorentz breaking physics and pushed well beyond expectations the tests of this fundamental symmetry of Nature, however still much seems to await along the path.

\begin{acknowledgement}
I wish to that Luca Maccione and David Mattingly for useful insights, discussions and feedback on the manuscript preparation.
\end{acknowledgement}

\bibliographystyle{spphys}	
\bibliography{references-LC}

\end{document}